\documentclass[useAMS,usenatbib,usegraphicx]{mn2e}
\usepackage[usenames, dvipsnames]{color}
\usepackage{subfigure}

\def\p1{\phantom{1}}

\def\simless{\mathbin{\lower 3pt\hbox
     {$\rlap{\raise 5pt\hbox{$\char'074$}}\mathchar"7218$}}}   
\def\simmore{\mathbin{\lower 3pt\hbox
     {$\rlap{\raise 5pt\hbox{$\char'076$}}\mathchar"7218$}}}   
\def\hide#1{}

\title[Phase lags of 4U 1636--53]{Phase lags of quasi-periodic oscillations across source states in the low-mass X-ray binary 4U 1636--53}
\author[M. G. B. de Avellar, M. M\'endez, D.Altamirano, A. Sanna \& G. Zhang]
{Marcio G. B. de Avellar$^{1}$\thanks{mgb.avellar@iag.usp.br}, 
Mariano M\'endez$^{2}$, 
Diego Altamirano$^{3}$, 
Andrea Sanna$^{4}$, 
\newauthor
Guobao Zhang$^{5}$\\ 
$^{1}$Instituto de Astronomia, Geof\'{i}sica e de Ci\^encias Atmosf\'ericas, Universidade de S\~ao Paulo, \\
Rua do Mat\~ao 1226, 05508-090, S\~ao Paulo, Brazil\\
$^{2}$Kapteyn Astronomical Institute, University of Groningen, P.O. Box 800, 9700 AV Groningen,\\
The Netherlands\\
$^{3}$Department of Physics and Astronomy, University of Southampton, Southampton, SO17 1BJ,\\
United Kingdom\\
$^{4}$Dipartimento di Fisica, Universit\`a degli Studi di Cagliari, SP Monserrato-Sestu km 0.7, I-09042, Monserrato,\\
Italy \\
$^{5}$New York University Abu Dhabi, P.O. Box 129188, \\
Abu Dhabi, United Arab Emirates}

\begin{document}

\date{
}

\pagerange{\pageref{firstpage}--\pageref{lastpage}} \pubyear{2013}

\maketitle

\label{firstpage}

\begin{abstract}

While there are many dynamical mechanisms and models that try to explain the origin and phenomenology of the quasi-periodic oscillations (QPOs) seen in the X-ray light curves of low-mass X-ray binaries, few of them address how the radiative processes occurring in these extreme environments give rise to the rich set of variability features actually observed in these light curves. A step towards this end comes from the study of the energy and frequency dependence of the phase lags of these QPOs. Here we used a methodology that allowed us to study, for the first time, the dependence of the phase lags of all QPOs in the range of 1 Hz to 1300 Hz detected in the low-mass X-ray binary 4U 1636--53 upon energy and frequency as the source changes its states as it moves through the colour-colour diagram. Our results suggest that within the context of models of up-scattering Comptonization, the phase lags dependencies upon frequency and energy can be used to extract size scales and physical conditions of the medium that produces the lags.

\end{abstract}


\clearpage

\section{Introduction}
\label{intro}

Neutron star and black hole low-mass X-ray binaries (NS- and BH-LMXBs) show rapid (seconds to milliseconds) X-ray variability components in the power density spectra (PDS). In the PDS of NS-LMXBs we can usually identify a power-law component at very low frequencies, a low-frequency complex of quasi-periodic oscillations (QPOs) ranging from $0.01$ to $100$ Hz, and a group of two to four QPOs with frequencies $\geq 100$ Hz \citep[see][for a review]{vanderklis01}.  

Most of the X-ray variability features identified as quasi-periodic oscillations have frequencies that correlate with each other in a given source while other properties of these QPOs, like the fractional amplitude (rms) and the quality factor (Q),  depend also on the spectral state of the source \cite[see, for example,][]{straaten02,diego01}. Hence, much of the work in the last 15 years has focused in the characterisation of the relation between the frequencies of different variability features or the dependence of the frequencies on other source properties \citep[e.g.][]{wijnands01,straaten02,disalvo01,mendez04,diego01}. In particular frequency-frequency correlations of some timing features across black-hole, neutron-star or white-dwarf X-ray binaries \citep[e.g.][]{wijnands02,psaltis01,mauche01,warner01} suggests that the origin of these features is independent of the nature of the central compact object, but they are produced in a physical component that is common to all these systems, e.g. the accretion disc or a corona. 

Of particular interest are the correlations between the fastest variability features, the so-called kiloHertz quasi-periodic oscillations (kHz QPOs), and other typical high- and low-frequency features \citep[e.g.][and references therein]{straaten02,straaten01,diego01}. Specially, the possibility that the frequency of the kHz QPOs trace the inner radius of the accretion disc would allow us to study the physics of matter under extreme regimes \citep[e.g.][]{miller01,stella01}.

Another aspect of the X-ray signal that has received renewed attention in the last few years is the energy- and frequency-dependent time (or phase) lags \citep[e.g.,][]{vanderklis02,wijers01,mitsuda01,vaughan01,kaaret01,nowak01,lee02,deAvellar01,barret06,kumar01}, since they can encode properties of the medium that produces the variability and the radiative processes occurring there.

Time/phase lags are Fourier-frequency-dependent measurements of the time (phase) delays between two concurrent and correlated signals, i.e. two light curves of the same source, in two different energy bands \citep[see][for more details on the subject]{nowak01}.

There are several papers describing observational results in a variety of systems over the years. The black hole system Cyg X-1, for example, displays hard lags ranging from 2 ms to several seconds with a $\sim \nu^{-0.7}$ frequency-dependence that increases logarithmically with energy \citep{miyamoto01,nowak01,bock01,grinberg01}.

Comparing BH- and NS-LMXBs, \cite{ford01} showed that in the frequency interval between 0.01 Hz and 100 Hz in both systems the lags are positive and in the range 0.03 to 0.2 rad.

\cite{mendez08} studied four galactic black hole systems reporting that the pair of high-frequency QPOs (HFQPOs) in GRS 1915+105 can be identified in broad terms to the pair of kHz QPOs in the neutron star systems 4U 1608--52 and 4U 1636--53. Both the lower kHz QPO in the neutron star systems and the lower HFQPO in GRS 1915+105 show soft lags that are inconsistent with the hard lags of the upper kHz QPO or the upper HFQPO, respectively. In the case of GRO J1655--40, the lags of the lower HFQPO are zero or slightly hard but significantly different from the hard lag of the upper HFQPO. At last, the similarity of the lag spectrum of the QPO detected at 180 Hz and after at 280 Hz in XTE 1550--564 lead them to conclude that they are the same QPO seen at different frequencies which, in turn, lead they to conclude that the lag spectrum could be used to identify the HFQPOs in these systems.

\cite{vaughan01}, \cite{kaaret01}, \cite{deAvellar01} and \cite{barret06} studied high frequency QPOs ($\geq 500$ Hz) in NS-LMXBs. They found soft lags for the lower kHz QPO with magnitudes weakly dependent upon frequency that increase with energy. \cite{deAvellar01} measured the lags of the upper kHz QPO showing that they are hard and that their magnitudes are independent of frequency and energy while inconsistent with the lags of the lower kHz QPO. 

The mechanisms proposed to explain these lags involve, in general, Compton up-/down-scattering of photons produced in the accretion disc (and in the case of neutron star systems, on the neutron-star surface or boundary layer) in a corona of hot electrons that surrounds the system \citep[see, for example,][for a discussion about the models]{sunyaev01,payne01,lee01,lee02,falanga01}.

Time lags are also found in the spectrum of gigantic systems like type 1 Seyfert galaxies and the active galaxy nuclei \citep[][for example]{zoghbi01,alston01}. In the case of active galactic nuclei, reflection of hard photons from the corona off the inner parts of the accretion disc \citep[e.g.][]{zoghbi01,zoghbi02} have been invoked to explain the observed soft lags. However, most of the stellar binary systems display hard lags that cannot be explained by the reflection model. Actually, \cite{cassatella01} have ruled out simple reflection models in order to explain these hard lags and although reflection off the accretion disc still could contribute to the lags, certainly it is not the dominant mechanism.

Here we present an extensive study of the time lags of all QPOs with frequencies above $1$ Hz detected in a large dataset of RXTE observations of the NS-LMXB 4U 1636--53 as a function of the position of the source on the colour-colour diagram (CCD). In \S2 we describe our observations and methodology; in \S3 we present the results for the frequency dependence of the phase lags (3.1) and for the energy dependence (3.2). We discuss our results and conclude in \S4.

\section{Observations and Data Analysis}
\label{obs}

From all available data of 4U 1636--53 taken with the PCA \citep{jahoda01} onboard the RXTE satellite \citep{bradt01} up to May 2010, we used only the 511 observations in which \cite{sanna02} detected kHz QPOs. 

For each of these observations we calculated the complex Fourier transform in nine energy bands, dividing each observation in contiguous 16-sec segments, and calculating the Fast Fourier Transform up to a Nyquist frequency of 2048 Hz. 

Because the time span of the dataset is about 15 years and in that period the gain of the instrument changed significantly,\footnote{See the channel-to-energy conversion table for the PCA at http://heasarc.gsfc.nasa.gov/docs/xte/e-c\_table.html} we adjusted our channel selections for different groups of observations, depending upon the epoch, in order to have approximately the same mean energy for each energy band at each epoch. We used in total seven narrow energy bands whose mean energies are 4.2 keV, 6.0 keV, 8.0 keV, 10.2 keV, 12.7 keV, 16.3 keV and 18.9 keV. These narrow energy bands comprise all the photons between approximately 3 keV to 5 keV, 5 keV to 7 keV, 7 keV to 9 keV, 9 keV to 11 keV, 11 keV to 15 keV, 15 keV to 17 keV and 17 keV to 20 keV, respectively\footnote{The exact edges of each band depend on the epoch.}. We also used two broad bands whose mean energies are 7.1 keV and 16.0 keV, the two broad bands comprising all photons with energies in the range 4 to 12 keV and all photons with energies between 12 and 20 keV, respectively. These are the same energy bands defined in \citet{deAvellar01}.

Prior to any calculation, we cleaned the X-ray light curves of all observations used in the present work from bursts, instrumental spikes and dropouts, since they would add power to the low frequency part of the PDS. We also subtracted the high-frequency average of the power spectra in the frequency range 1300 to 2000 Hz. Dead-time corrections were made to get the colours (see below) as explained in \cite{guobao01}.

We calculated the phase lags of the QPOs following \citet{vaughan01} and \citet{kaaret01}. 

Under the assumption that the PDS properties correlate with the source position in the CCD, i.e., the PDS shows similar shapes and features depending on the position, \citep{straaten02,straaten01,diego01}, we relied on the spectral states of the source to track the variability features more accurately. For this we first calculated the colours of our observations using RXTE's Standard 2 data. We defined the  9.7-16.0 keV/6.0-9.7 keV count rate ratio as the hard colour (HC) and the 3.5-6.0 keV/2.0-3.5 keV count rate ratio as the soft colour (SC). To  correct for gain changes and differences in the effective area between the proportional counter units (PCUs)  we normalized our  colours to those of the Crab nebula obtained from observations that were close in time to our own observations. Finally we averaged the normalized colours per PCU for the full observation using all available PCUs. The details of this procedure can be found in \cite{kuulkers01} and \cite{diego01}.

Then, to study the phase lags of all QPOs on the assumption that the properties of the PDS correlate with the source position in the CCD, we defined boxes in the CCD in the following way:

\begin{itemize}

\item the boxes should be small to avoid changes in the physical conditions of the source (which would imply changes in the shape of the PDSs, specially in the bottom part of the CCD where the source is more variable) but, at the same time, the number of observations within each box should be large enough to have the best statistics possible to build the average spectra;

\item after a first definition of 27 boxes, we compared the shapes of the PDSs of each individual observation within each box both, visually and statistically\footnote{By statistically we mean that we analysed the average of two probabilities of similarity, KS \citep{kolgomorov01} and MTT \citep{melo01}, and we say that two PDSs are similar if the average of the probabilities are greater than 65\%.}. If we identified, within a box, groups of observations with similar PDSs, then we further subdivided the box by grouping these similar observations.

\end{itemize}

\begin{figure}
\centering
{
     \includegraphics[width=0.420 \textwidth, angle=0]{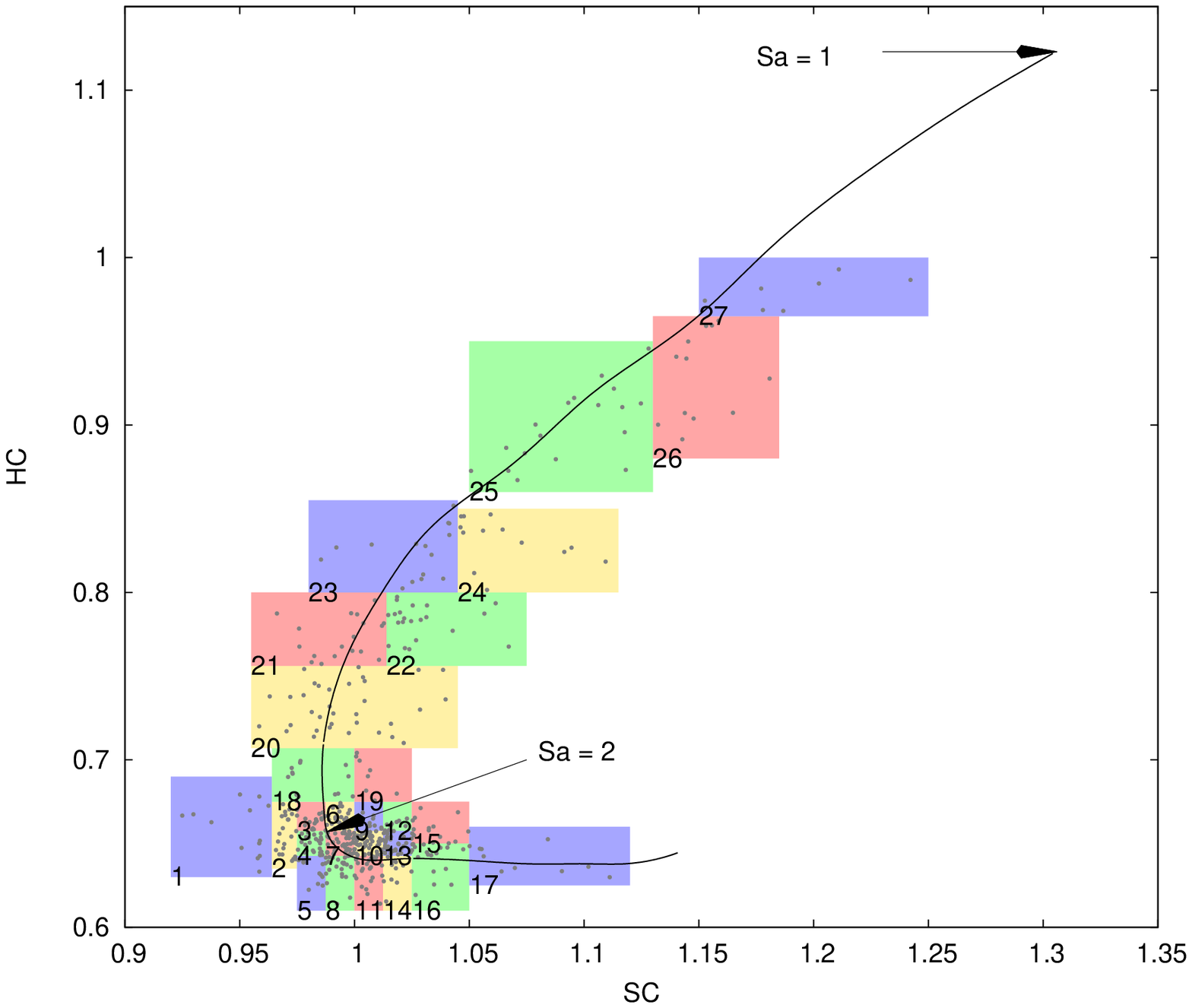}
     \includegraphics[width=0.450 \textwidth, angle=0]{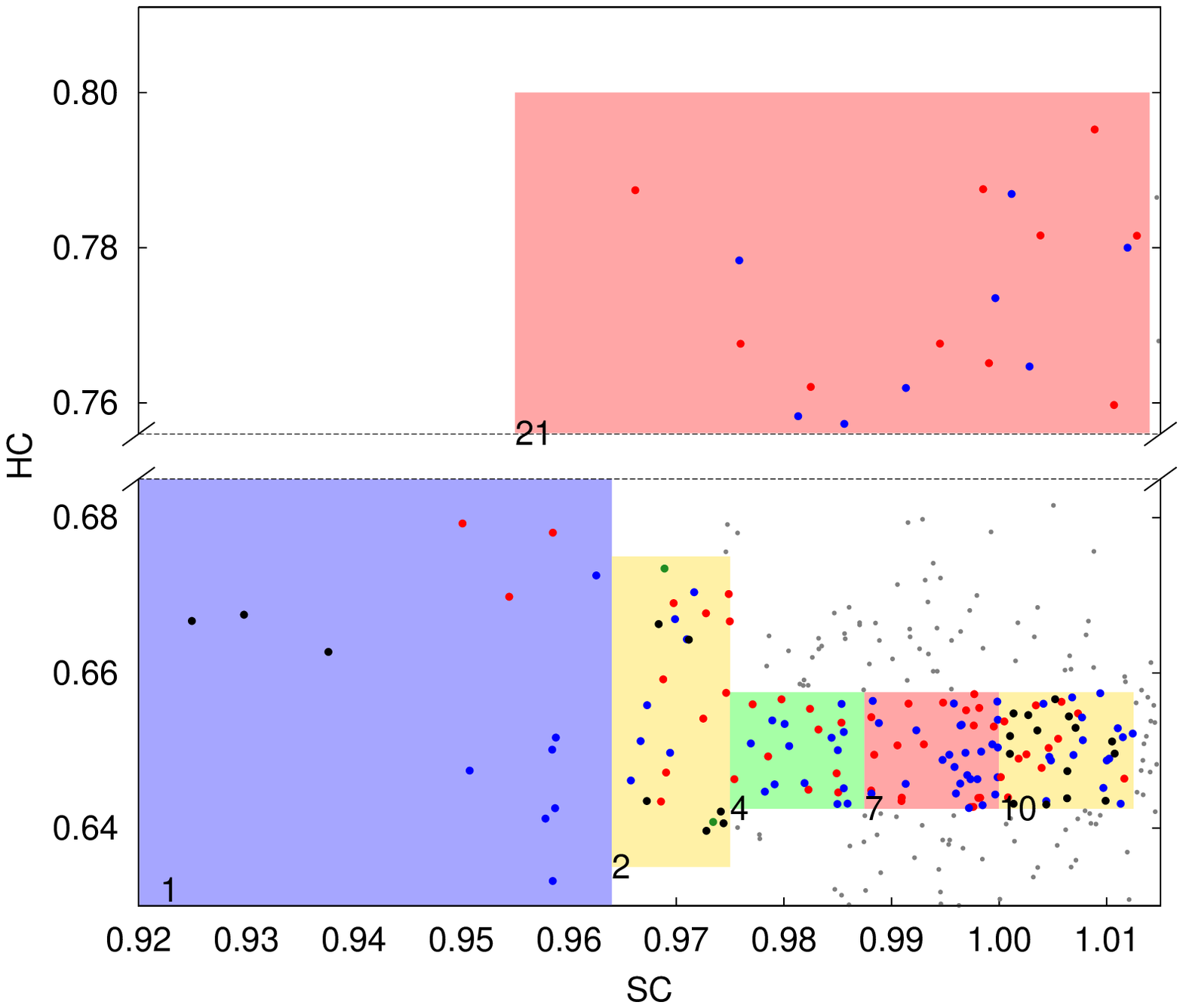}
}

\caption{In the upper panel we show the colour-colour diagram of 4U 1636--53 showing the 27 originally defined boxes. In the lower panel we show only the boxes that were further subdivided: each of these boxes displays groups of observations (highlighted by different colour points) with similar PDSs. Thus, for a given box, we grouped the points with same colour creating the subdivisions. Each point corresponds to a single RXTE observation. We only include in this diagram the 511 observations with kHz QPOs (see text). For each box we fitted the features displayed in the average power density spectrum. See Figure \ref{fig:averageSpectraAndFeatures}.}

\label{fig:CCDboxes}
\end{figure}

We ended up with 37 boxes comprising 2 to 34 observations, as depicted in Figure \ref{fig:CCDboxes}.

%
%
%

We fitted the features ranging from 1 Hz to 1300 Hz seen in the average power density spectrum of each box with Lorentzian functions, which depend on three parameters: the integral power, the central frequency ($\nu_{0}$) and the full-width at half maximum (FWHM). This procedure provided us with the properties of the QPOs that appear in each given box. In some cases, for example box 19 , we used a non-QPO component to better fit the PDS at low frequencies $\leq 1$ Hz (shown as a dashed line with no label in Figure \ref{fig:averageSpectraAndFeatures}). This non-QPO component is a Lorentzian with centroid frequency fixed at zero. We parametrised the position on the CCD with the quantity $S_{a}$ \citep{guobao01,mendez03} which is derived from the parametrization $S_{Z}$ \citep[and references therein]{wijnands03}. In this phenomenological approach we approximated the shape traced by the observations on the CCD with a spline where the quantity $S_{a}$ along the the spline is a measure of the position of the source. See Figure \ref{fig:CCDboxes}. Here we fixed $S_{a}=1$ at (SC,HC) = (1.31,1.12) and $S_{a}=2$ at (SC,HC) = (0.988,0.657).

\begin{figure}
\centering
{
     \includegraphics[width=0.480 \textwidth, angle=0]{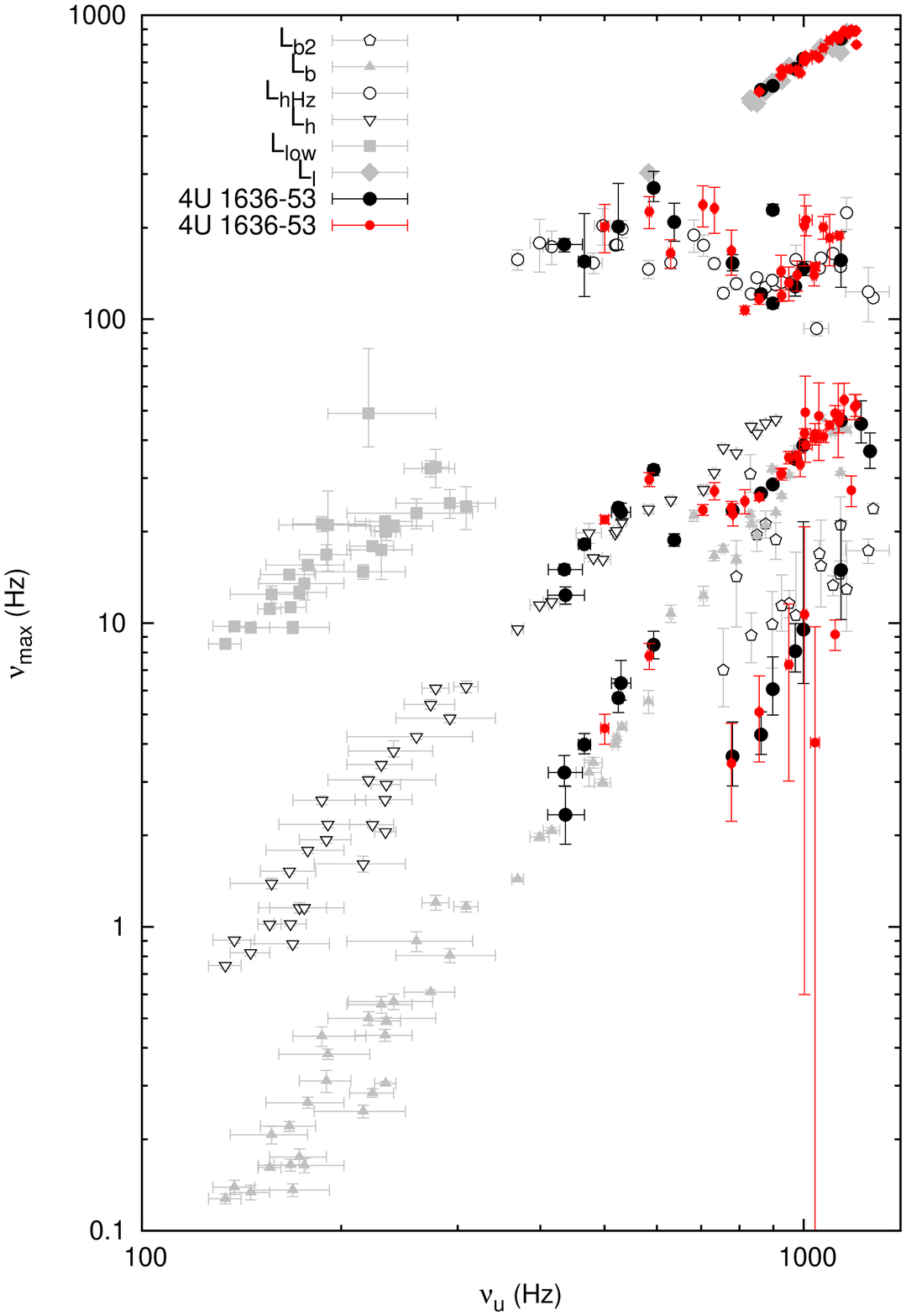}
}

\caption{Correlations between the characteristic frequencies $\nu_{max}$ of the $L_{b2}$, $L_{b}$, $L_{h}$, $L_{hHz}$, $L_{l}$ QPOs and the characteristic frequency $\nu_{u}$ of $L_{u}$. Symbols other than the black and red points represent the atoll sources 4U 0614+09, 4U 1728--34 \citep{straaten02}, 4U 1608--52 \citep{straaten01} and Aql X-1 \citep{reig01}. Also included are the low-luminosity bursters  1E 1724--3045, GS 1826--24, and SLX 1735--26 \citep{straaten03}. Black dots are 4U 1636--53 data from \citealt{diego01}. Red dots are the 4U 1636--53 present data. Figure adapted from \citealt{diego01}.}
\label{fig:diagnostico1}
\end{figure}

\begin{figure}
\centering
\subfigure[RMS amplitude of the six QPOs of 4U 1636--53 against the frequency of the upper kiloHertz QPO. Red dots were calculated by setting the background counts to zero, while the blue dots were calculated by setting the background counts to a maximum average for the PCAs.] 
{
     \label{rms-nu}
    \includegraphics[width=0.480 \textwidth, angle=0]{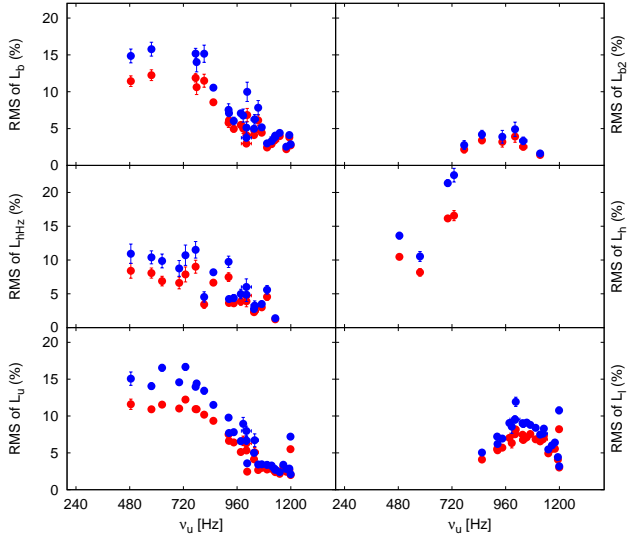}
}
\hspace{1cm}
\subfigure[Quality factor of the six QPOs of 4U 1636--53 against the frequency of the upper kiloHertz QPO.] 
{
     \label{Q-nu}
    \includegraphics[width=0.480 \textwidth, angle=0]{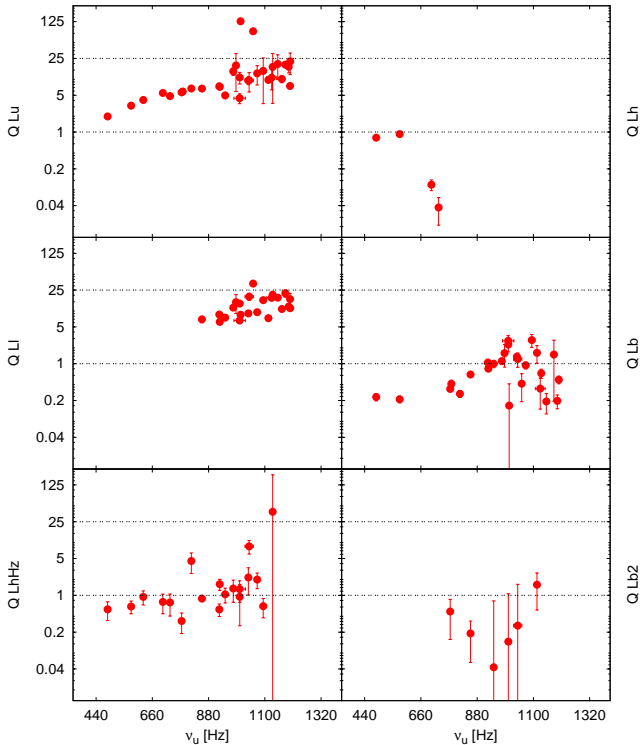}
}
\caption{The results rms-$\nu_{u}$ and Q-$\nu_{u}$ are in agreement with previous works \citep[see][for example]{diego01}. All data were obtained with the presented methodology.} 
\label{fig:diagnostico2}
\end{figure}

\begin{figure}
\centering
{
     \includegraphics[width=0.480 \textwidth, angle=0]{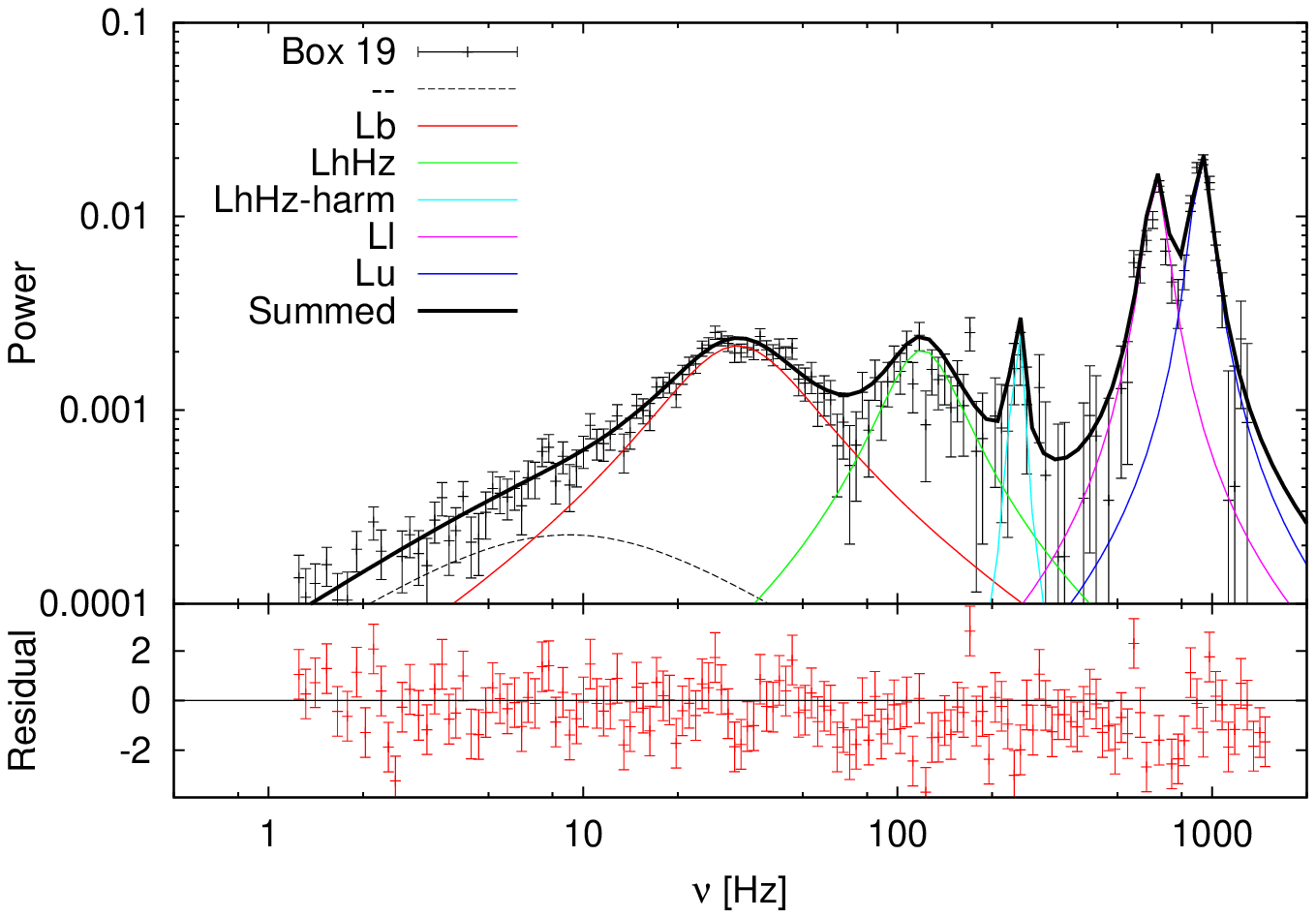}
     \includegraphics[width=0.480 \textwidth, angle=0]{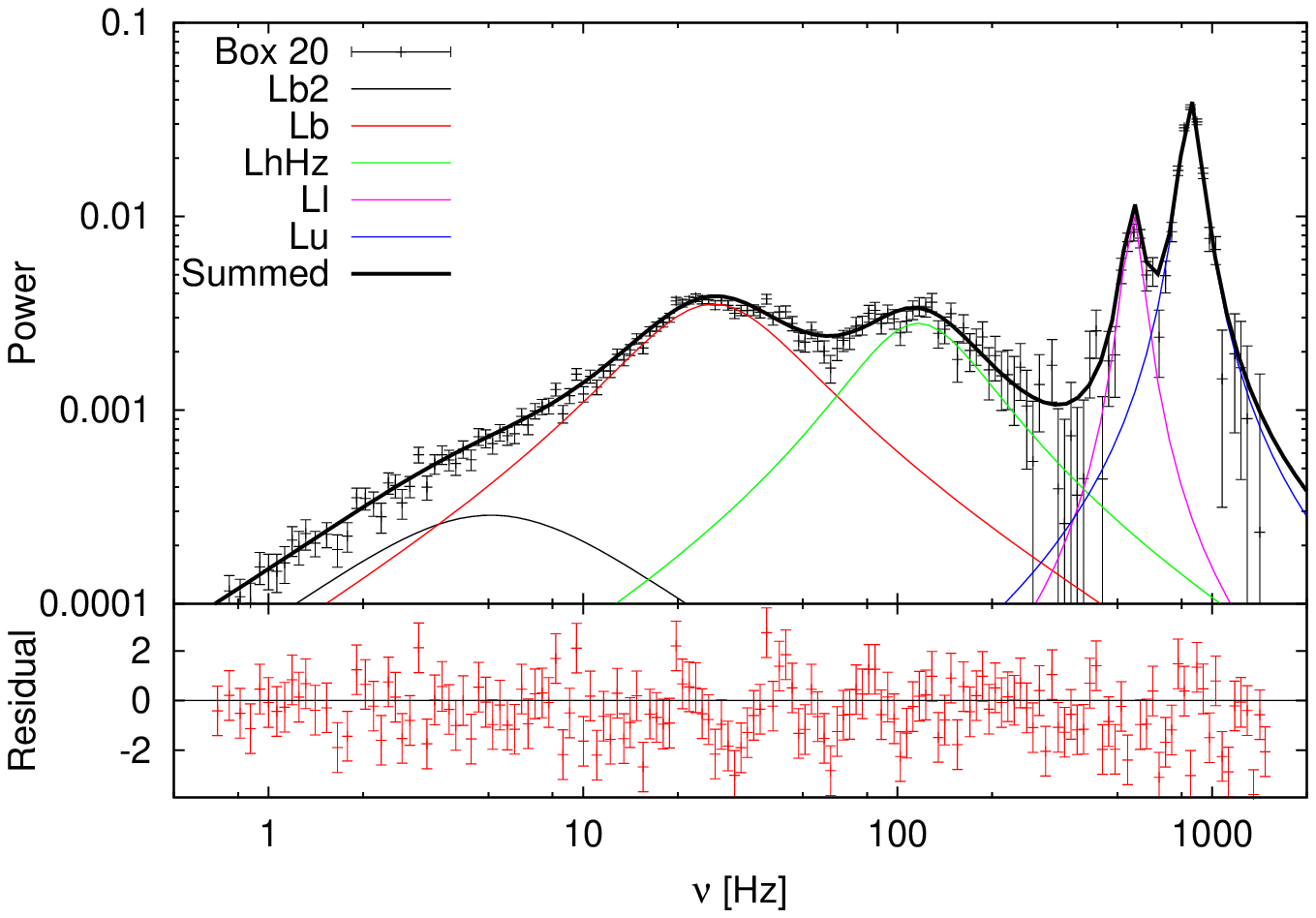}
     \includegraphics[width=0.480 \textwidth, angle=0]{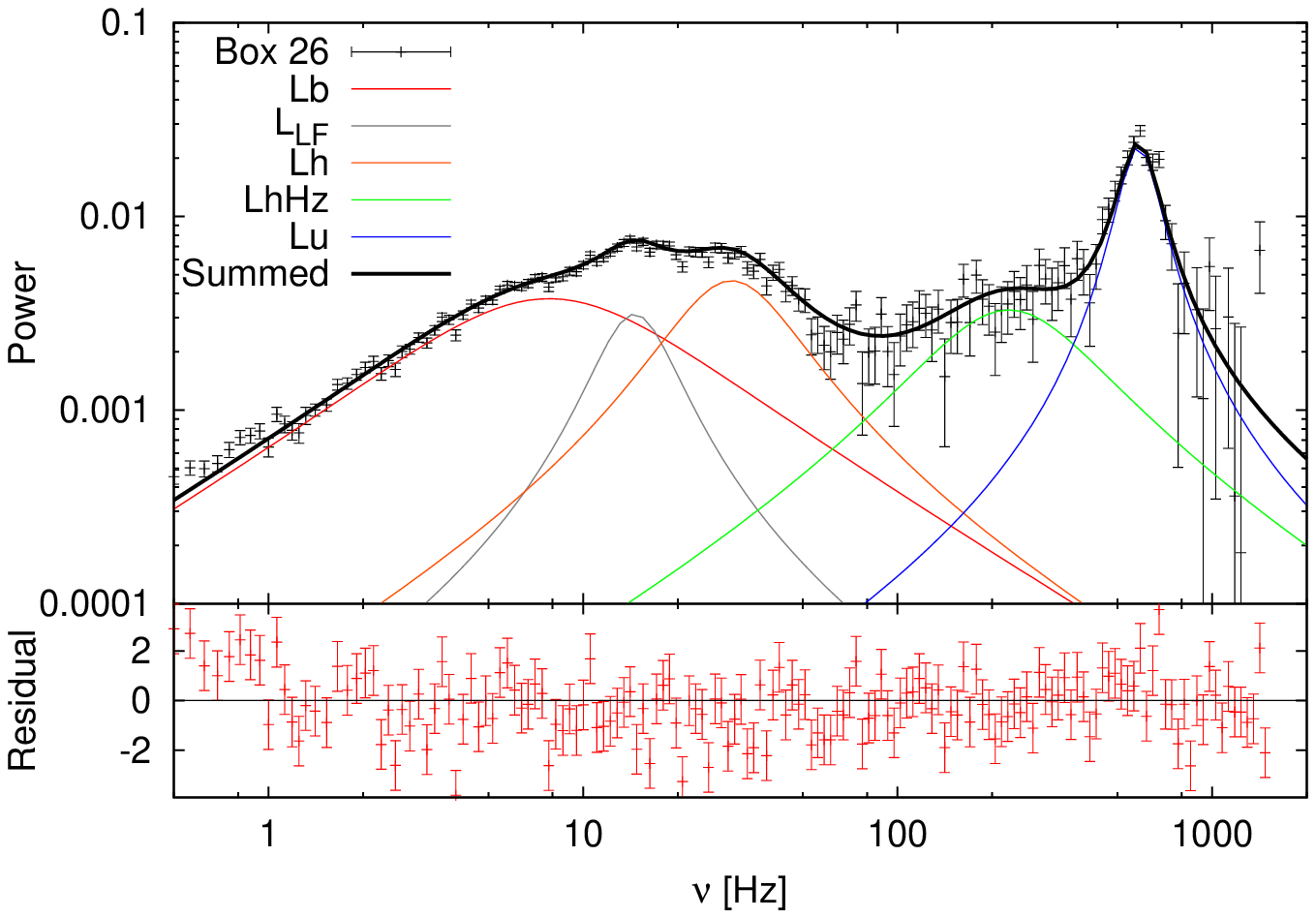}
}

\caption{Average power density spectra of three selected boxes in the CCD of 4U 1636--36 showing the fitting components. The labels are: $L_{b2}$ for the QPO where the PDS shows a second break, $L_{b}$ for the QPO where the PDS breaks, $L_{LF}$ for the low frequency QPO, $L_{h}$ for the ``hump'' QPO, $L_{hHz}$ for the hectoHertz QPO, $L_{hHz-harm}$ for the first harmonic of the hectoHertz QPO, $L_{l}$ for the lower kiloHertz QPO and $L_{u}$ for the upper kiloHertz QPO. These three boxes were chosen as examples because each displays many QPOs and together they sample all the eight different QPOs we detect through the data, including $L_{LF}$ and $L_{hHz-harm}$ that we did not study here. Notice that the dashed feature in box 19 is not a QPO: it is a lorentzian with centroid frequency fixed at zero used to better fit the PDS at low frequencies $\leq 1$ Hz. We also show in the panels the summed up models and the residuals of the fits.}
\label{fig:averageSpectraAndFeatures}
\end{figure}

Note that not all QPOs are present, or significantly detected, in all boxes: the appearance of a given QPO depends on the position of the source in the CCD, as mentioned before. Then the question arises of how secure are the QPO identifications in the average spectrum of each box. To identify the lower and upper kiloHertz QPOs we used the well known relation of the (centroid) frequency of these two QPOs with the hard colour (see Figure 1 of \citealt{sanna02} and Figure 3 of \citealt{belloni01}). To identify the other QPOs more precisely we relied on the correlations of their frequency, rms and quality factor with the frequency of the upper kHz QPO calculated for our data and compared with \cite{diego01} (see Figures \ref{fig:diagnostico1} and \ref{fig:diagnostico2} here and Figures 5, 10, 11 and 12 of \citealt{diego01}). In six cases where a kHz QPO in the range 800 to 900 Hz appeared alone we classified them as the lower kHz QPO since Q vs frequency and rms vs frequency did not obeyed the expected relation of the upper kHz QPO and the colour was compatible with the other cases where the lower kHz QPO identification was unambiguous. The criteria we used to assess whether a  QPO is significantly detected is if the ratio between the integral power and its uncertainty is larger than 3, but allowing to be $2.5 < \sigma < 3$ if the QPO lies in all the correlations mentioned above. In total, we detected eight different QPOs that, following \cite{diego01}, we called $L_{b2}$, $L_{b}$, $L_{LF}$, $L_{h}$, $L_{hHz}$, $L_{hHz-harm}$, $L_{l}$, $L_{u}$.

In Figure \ref{fig:averageSpectraAndFeatures} we show three boxes as examples of average spectra. The three boxes were chosen because they show many of the QPOs and these three boxes together sample all the eight different QPOs we detected through the data, including $L_{LF}$ and $L_{hHz-harm}$ that we did not study in this work. In Table \ref{tab:QPOsEachBox} we show the QPOs detected in each box.
%
%
%

%
%
%

%
%
%

%
%
%
\begin{table}
\caption{Detected QPOs of the NS-LMXB 4U 1636--53 through the colour-colour diagram. Although we detected eight different QPOs, $L_{LF}$ and $L_{hHz_harm}$ are not shown because they appeared and only two occasions and were not studied here.}
\centering
\begin{tabular}{|c|c|c|c|c|c|c|}
\hline
Box & \multicolumn{6}{|c|}{Detected QPOs} \\
\hline
    Box 1-1 &  & $L_{b}$ &  &  & $L_{l}$ & $L_{u}$ \\
    Box 1-2 &  & $L_{b}$ &  &  & $L_{l}$ & $L_{u}$ \\
    Box 1-3 &  &  &  &  & $L_{l}$ &  \\
    Box 2-1 &  & $L_{b}$ &  & $L_{hHz}$  & $L_{l}$ & $L_{u}$ \\
    Box 2-2 &  & $L_{b}$ &  &  & $L_{l}$ & $L_{u}$ \\
    Box 2-3 &  &  &  &  & $L_{l}$ & $L_{u}$ \\    
    Box 2-4 &  &  &  & $L_{hHz}$ & $L_{l}$ &  \\
    Box 3 &  & $L_{b}$ &  & $L_{hHz}$ & $L_{l}$ & $L_{u}$ \\
    Box 4-1 & $L_{b2}$ & $L_{b}$ &  & $L_{hHz}$ & $L_{l}$ & $L_{u}$ \\
    Box 4-2 &  &  & $L_{h}$ &  & $L_{l}$ &  \\
    Box 5 & $L_{b2}$ & $L_{b}$ &  & $L_{hHz}$ & $L_{l}$ & $L_{u}$ \\
    Box 6 & $L_{b2}$ & $L_{b}$ &  & $L_{hHz}$ & $L_{l}$ & $L_{u}$ \\
    Box 7-1 &  & $L_{b}$ &  & $L_{hHz}$ & $L_{l}$ & $L_{u}$ \\
    Box 7-2 &  & $L_{b}$ &  & $L_{hHz}$ & $L_{l}$ & $L_{u}$ \\
    Box 8 & $L_{b2}$ &  & $L_{h}$ &  & $L_{l}$ & \\
    Box 9 & $L_{b2}$ & $L_{b}$ &  &  & $L_{l}$ & $L_{u}$ \\
    Box 10-1 &  & $L_{b}$ &  & $L_{hHz}$ & $L_{l}$ & $L_{u}$ \\
    Box 10-2 &  & $L_{b}$ &  & $L_{hHz}$ & $L_{l}$ & $L_{u}$ \\
    Box 10-3 &  & $L_{b}$ &  &  & $L_{l}$ & $L_{u}$ \\
    Box 11 &  &  & $L_{h}$ &  & $L_{l}$ & \\
    Box 12 &  & $L_{b}$ &  &  & $L_{l}$ & $L_{u}$ \\
    Box 13 &  & $L_{b}$ &  &  & $L_{l}$ & $L_{u}$ \\
    Box 14 &  &  &  &  & $L_{l}$ & \\
    Box 15 &  &  &  &  & $L_{l}$ & $L_{u}$ \\
    Box 16 &  & $L_{b}$ &  &  & $L_{l}$ & $L_{u}$ \\
    Box 17 &  & $L_{b}$ &  &  & $L_{l}$ & $L_{u}$ \\
    Box 18 &  & $L_{b}$ &  & $L_{hHz}$ & $L_{l}$ & $L_{u}$ \\
    Box 19 &  & $L_{b}$ &  & $L_{hHz}$ & $L_{l}$ & $L_{u}$ \\
    Box 20 & $L_{b2}$ & $L_{b}$ &  & $L_{hHz}$ & $L_{l}$ & $L_{u}$ \\
    Box 21-1 &  & $L_{b}$ &  &  &  & $L_{u}$ \\
    Box 21-2 &  & $L_{b}$ &  & $L_{hHz}$ &  & $L_{u}$ \\
    Box 22 & $L_{b2}$ & $L_{b}$ &  & $L_{hHz}$ &  & $L_{u}$ \\
    Box 23 &  &  & $L_{h}$ & $L_{hHz}$ &  & $L_{u}$ \\
    Box 24 &  &  & $L_{h}$ & $L_{hHz}$ &  & $L_{u}$ \\
    Box 25 &  &  &  & $L_{hHz}$ &  & $L_{u}$ \\
    Box 26 &  & $L_{b}$ & $L_{h}$ & $L_{hHz}$ &  & $L_{u}$ \\
    Box 27 &  & $L_{b}$ & $L_{h}$ & $L_{hHz}$ &  & $L_{u}$ \\
\hline
\hline
\end{tabular}
\label{tab:QPOsEachBox}
\end{table}

Phase lags are relative quantities, i.e., one calculates the lags of the photons of one energy band relatively to the photons of another energy band (the bands defined arbitrarily). We therefore defined {\it soft}, or {\it negative}, lags when the photons of the less energetic band lag behind the photons of the more energetic band. On the other hand, when the photons of the more energetic band lag behind the photons of the less energetic band we say that the lags are {\it hard}, or {\it positive}.

Here, we studied the frequency and energy dependence of the phase lags of the photons of all the detected QPOs in each box. For the energy-dependence of the lags, we calculated the phase lags of the photons of each narrow energy band relatively to the 10.2 keV band\footnote{We chose the 10.2 keV band because in this band both the variability and the source intensity are high, which reduces the uncertainties of the of time-lag measurements \citep{vaughan03}.}. For the frequency-dependence of the lags, we calculated the phase lags of all photons in the broad band above 12 keV relatively to all photons in the broad band below 12 keV.

\begin{figure}
\centering
{
     \includegraphics[width=0.320 \textwidth, angle=-90]{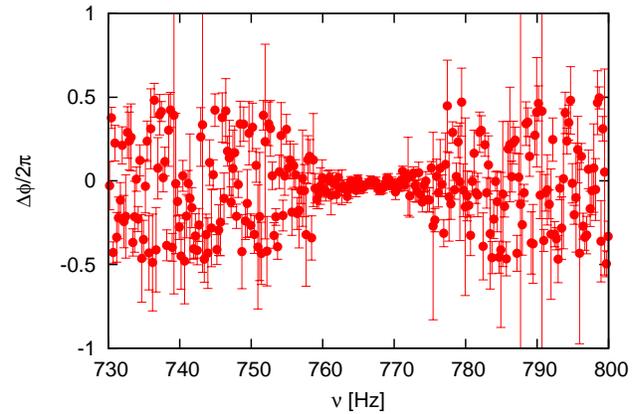}
}

\caption{Phase lag as a function of Fourier frequency in 4U 1636--53, in the frequency range around a lower kiloHertz QPO whose centroid frequency is at $\simeq 766$ Hz in a given moment. Notice that we averaged the phase lags over one FWHM interval which, in this case, is apparent between 760 and 775 Hz.}
\label{fig:lagSpectraIlustration}
\end{figure}

Finally, to calculate the lags of the QPOs we averaged the energy- and frequency-dependent lags over a frequency interval of one FWHM around the centroid frequency of the QPO, except in the cases where two QPOs are broad enough to overlap each other. In these cases we averaged over a fraction of the FWHM ranging from one half to one fifth. The price paid was to admit larger error bars in the average lag due to less data points in the respective interval. Unless otherwise noted, all phase lags are given in units of $2\pi$ rad. In Figure \ref{fig:lagSpectraIlustration} we show an example of the lag spectra of the lower kHz QPO in a given moment to illustrate the interval over which we averaged the lags.

%
%
%

A final remark about the dataset we used in this work should be made. We selected only the 511 observations in which \cite{sanna02} detected at least one kHz QPO and therefore we assume that because the position of the source in the CCD determines, at least to some extent, the shape of the PDS and the components that appear, the characteristics of the other components (i.e., $L_{b2}$, $L_{b}$, $L_{h}$, $L_{hHz}$, that correlate with the upper kHz QPO \citep{diego01}) will not change in observations that do not show kHz QPOs.

\section{Results}
\label{res}

\subsection{Frequency dependence of the phase lags}
\label{res1}

\begin{figure*}
\centering
{
    \includegraphics[width=0.325 \textwidth,angle=-90]{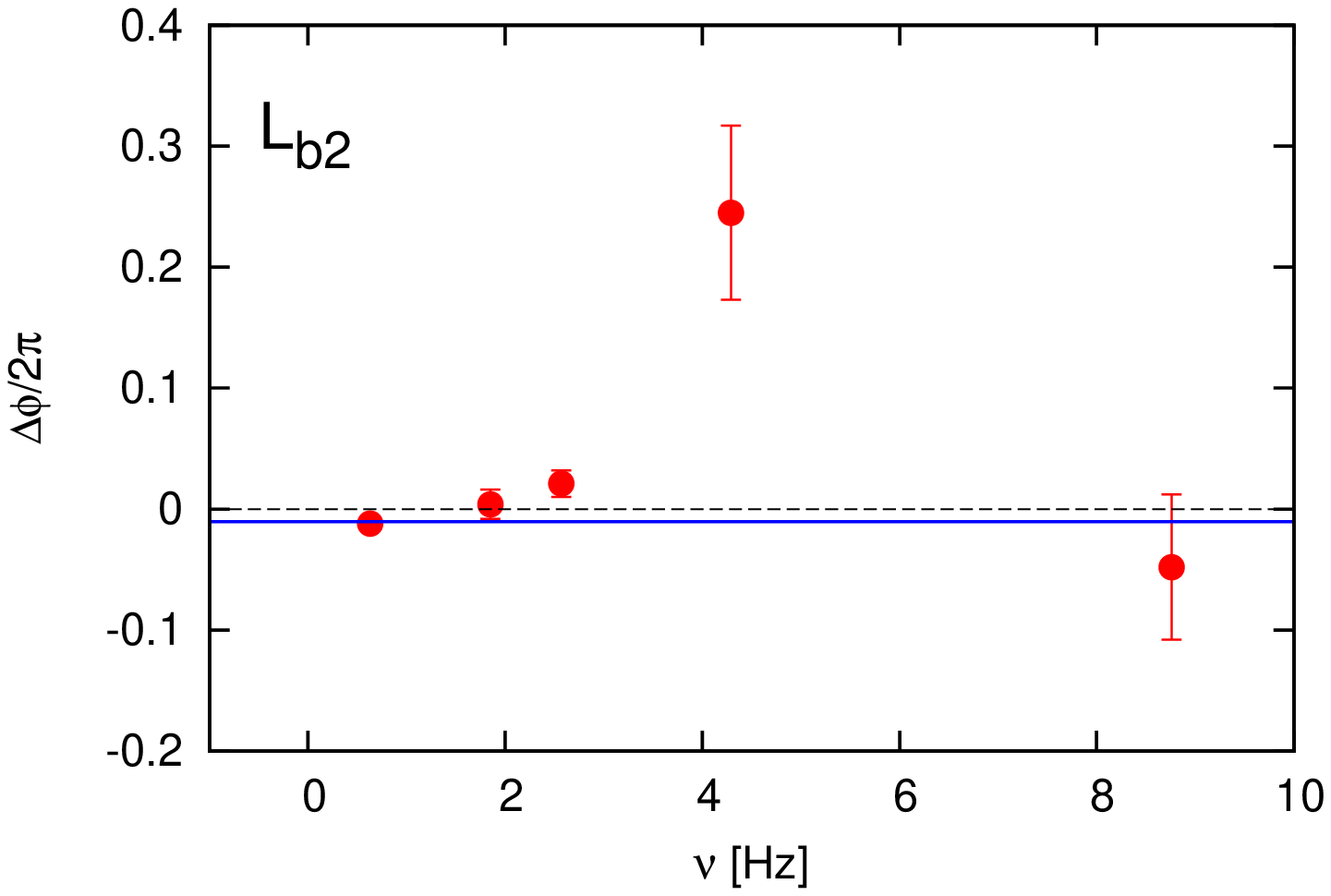}
}
{
    \includegraphics[width=0.325 \textwidth,angle=-90]{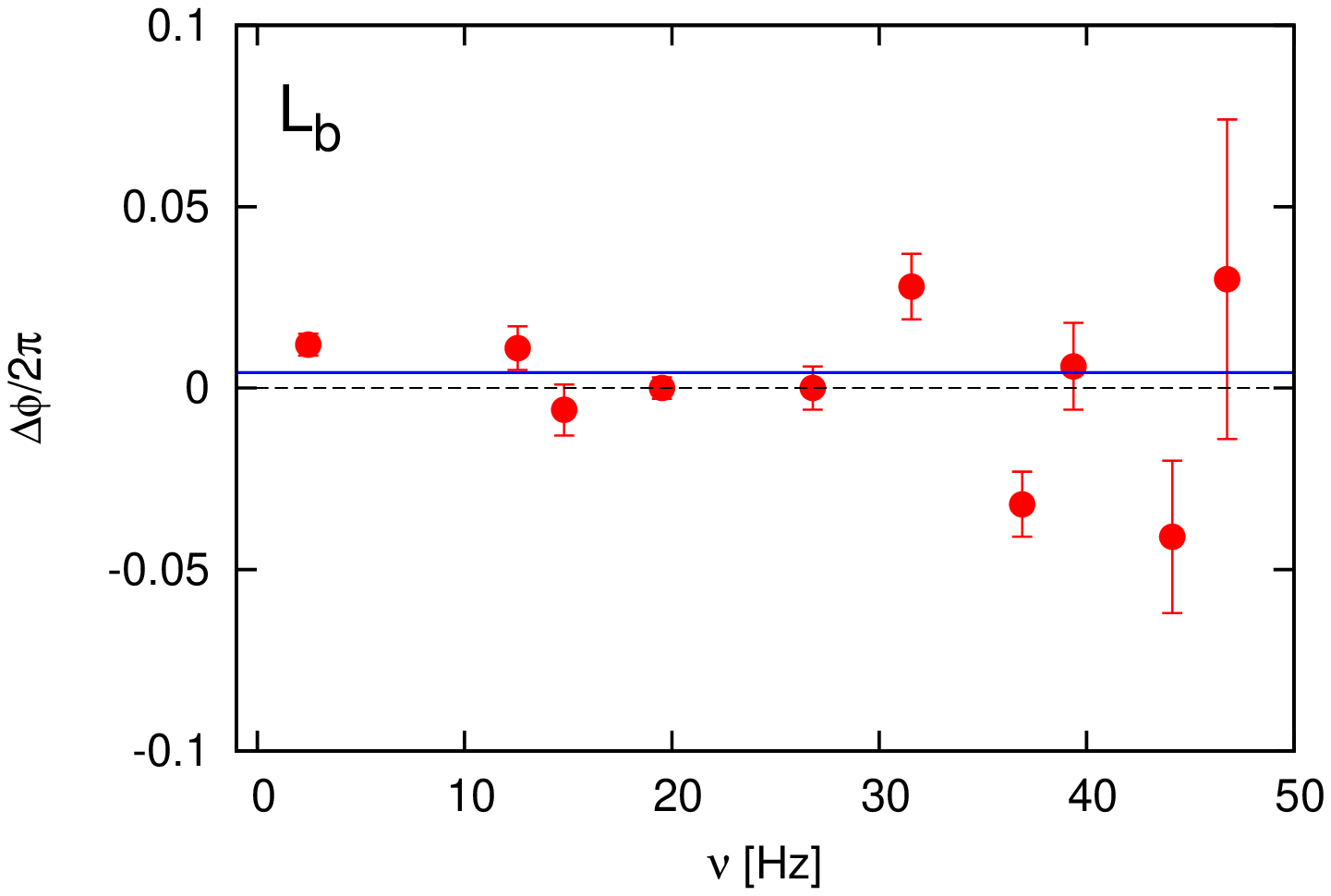}
}
{
    \includegraphics[width=0.325 \textwidth,angle=-90]{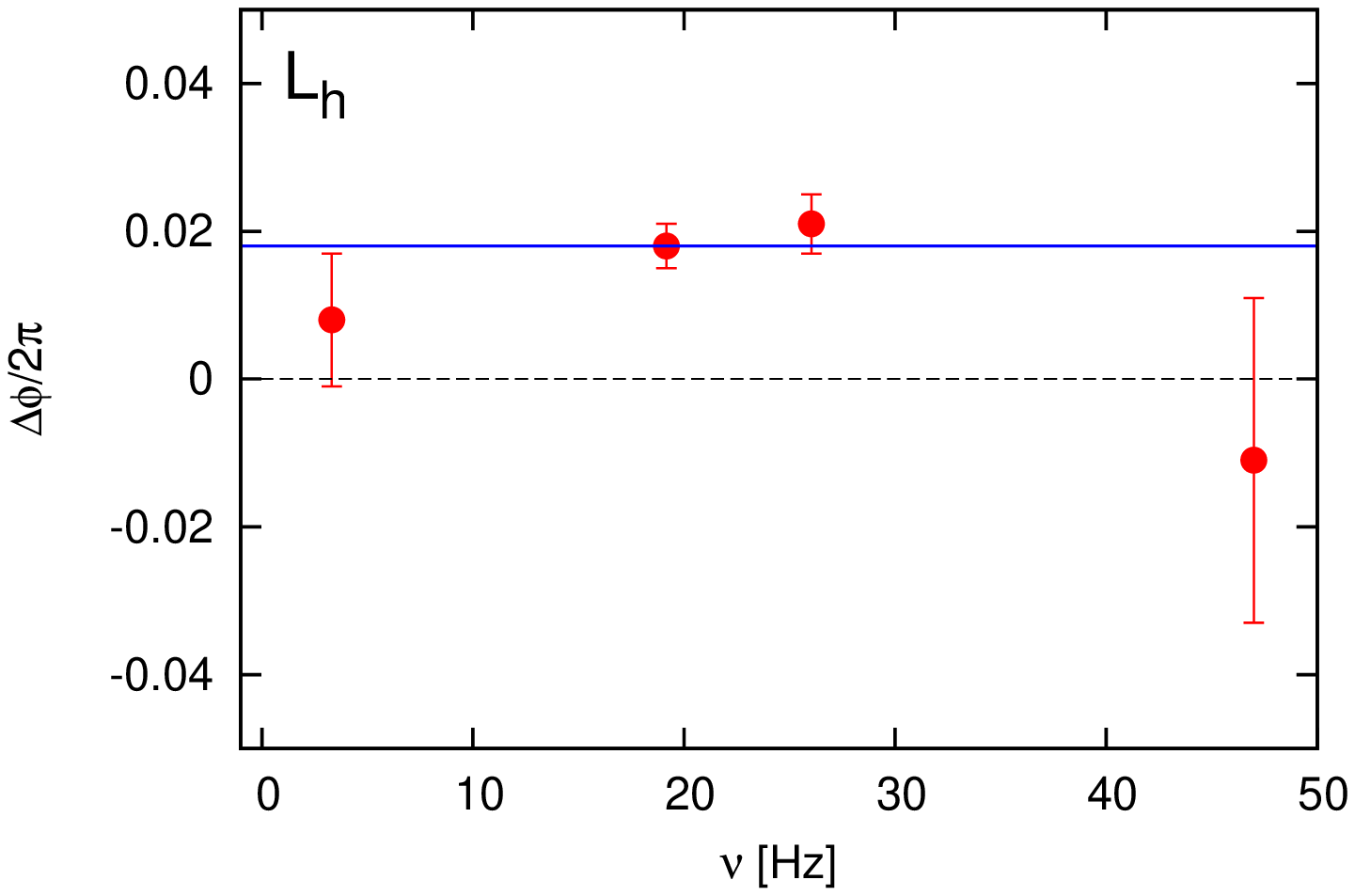}
}
{
    \includegraphics[width=0.325 \textwidth,angle=-90]{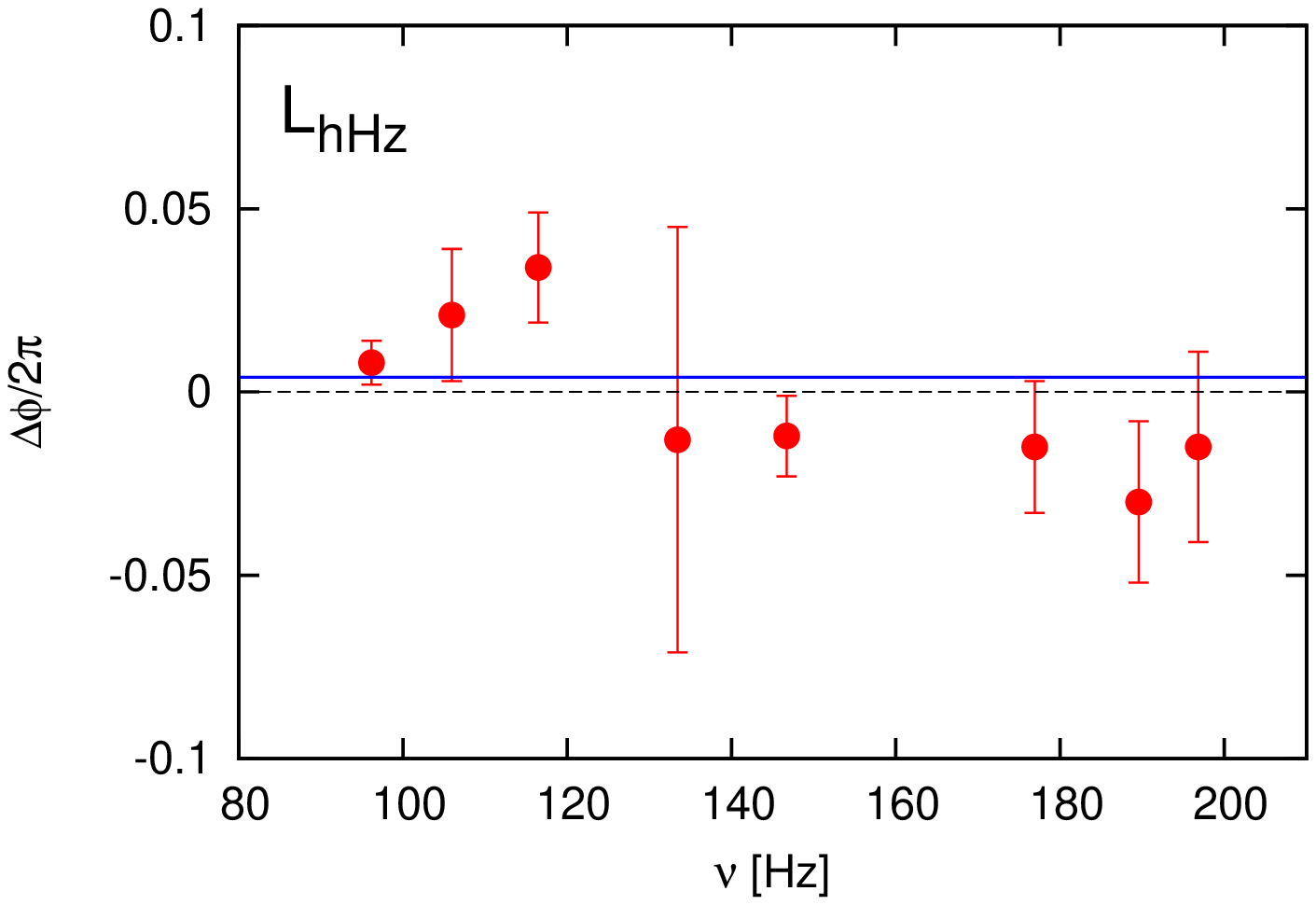}
}
{
    \includegraphics[width=0.325 \textwidth,angle=-90]{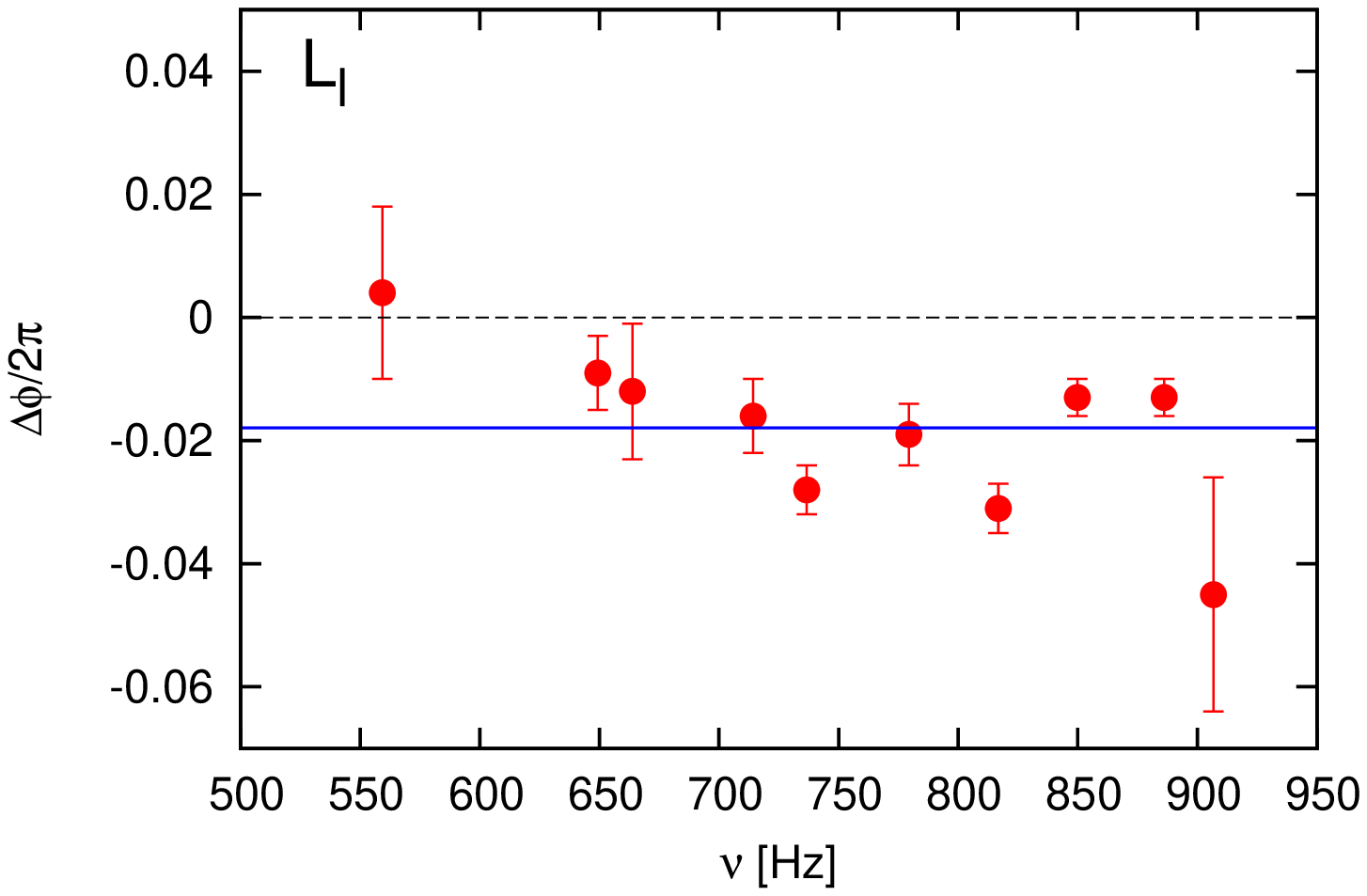}
}
{
    \includegraphics[width=0.325 \textwidth,angle=-90]{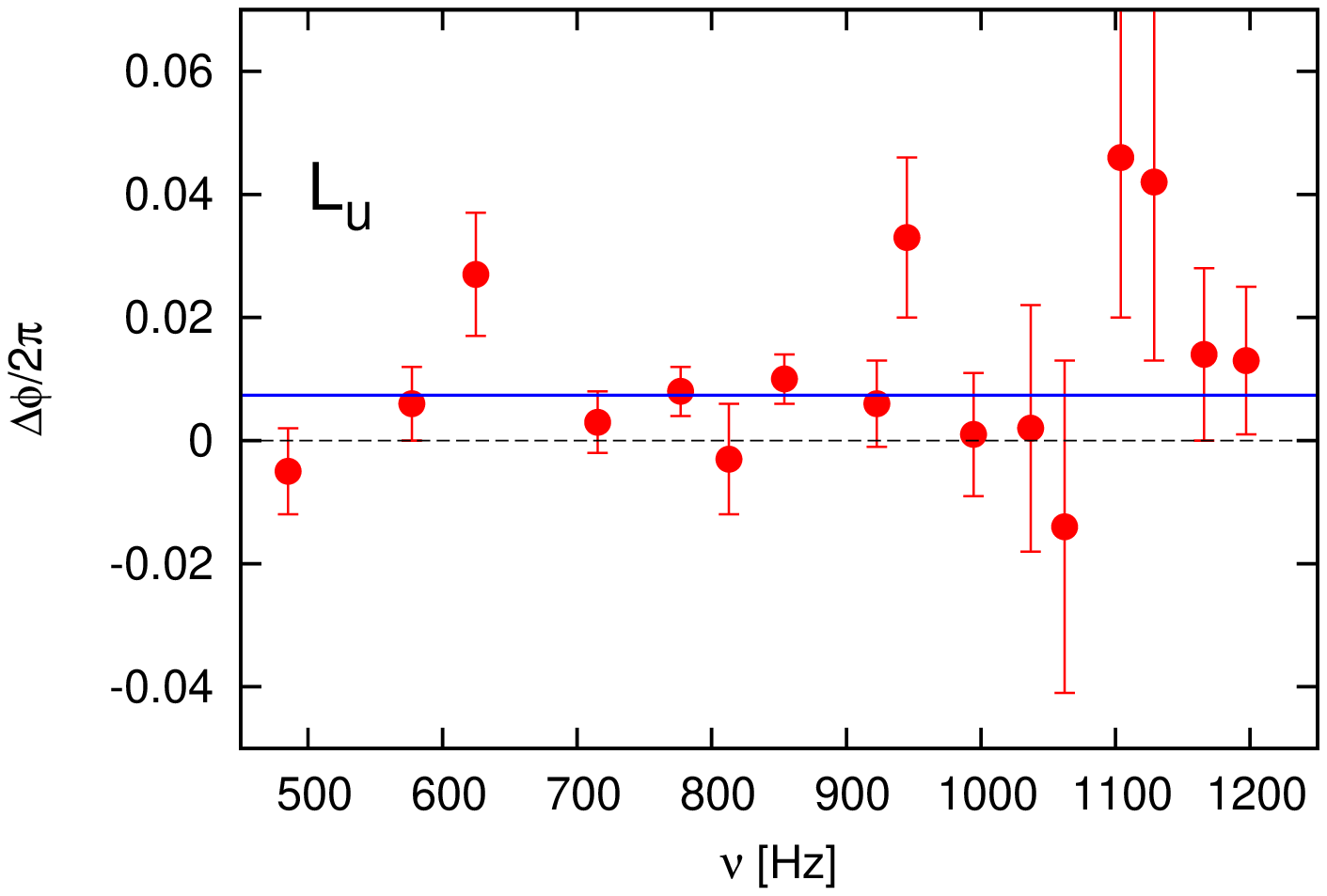}
}
\caption{Phase lags, in unit of $2\pi$, of the different QPO components of 4U 1636--53 as a function of the frequency of the corresponding QPO. The lags represent the delay of photons with energies in the 12--20 keV range (mean energy $\simeq 16$ keV) relative to the photons with energies in the 4--12 keV range (mean energy $\simeq 7$ keV). Left top panel: the QPO where the PDS shows a second break. Right top panel: the QPO at the break frequency. Left middle panel: the {\it hump} QPO. Right middle panel: the hecto-hertz (hHz) QPO. Left bottom panel: the lower kHz QPO. Right bottom panel: the upper kHz QPO. We did not include the lags for the $L_{LF}$ ($\sim 0.01$ at 10 Hz) and $L_{hHz-harm}$ ($\sim 0.1$ at 240 Hz) QPOs since we have only two and one points, respectively. The blue lines are the best fit constant to the lags. The errors correspond to the $1\sigma$ confidence level.}
\label{fig:phaseLagsVSfrequency} 
\end{figure*}

In Figure \ref{fig:phaseLagsVSfrequency} we plot the phase lags of six out of the eight QPOs detected in the PDS as a function of the frequency of the QPO for photons with energies in the band 12-20 keV (mean energy $\simeq 16.0$ keV) relatively to photons with energies in the band 4-12 keV (mean energy $\simeq 7.1$ keV). We do not show results for $L_{LF}$ and $L_{hHz-harm}$ since we detected $L_{LF}$ in two occasions and $L_{hHz-harm}$ in only one occasion. Uncertainties are given at a 68\% confidence level.  

%
%
%

\begin{figure*}
\centering
{
    \includegraphics[width=0.325 \textwidth,angle=-90]{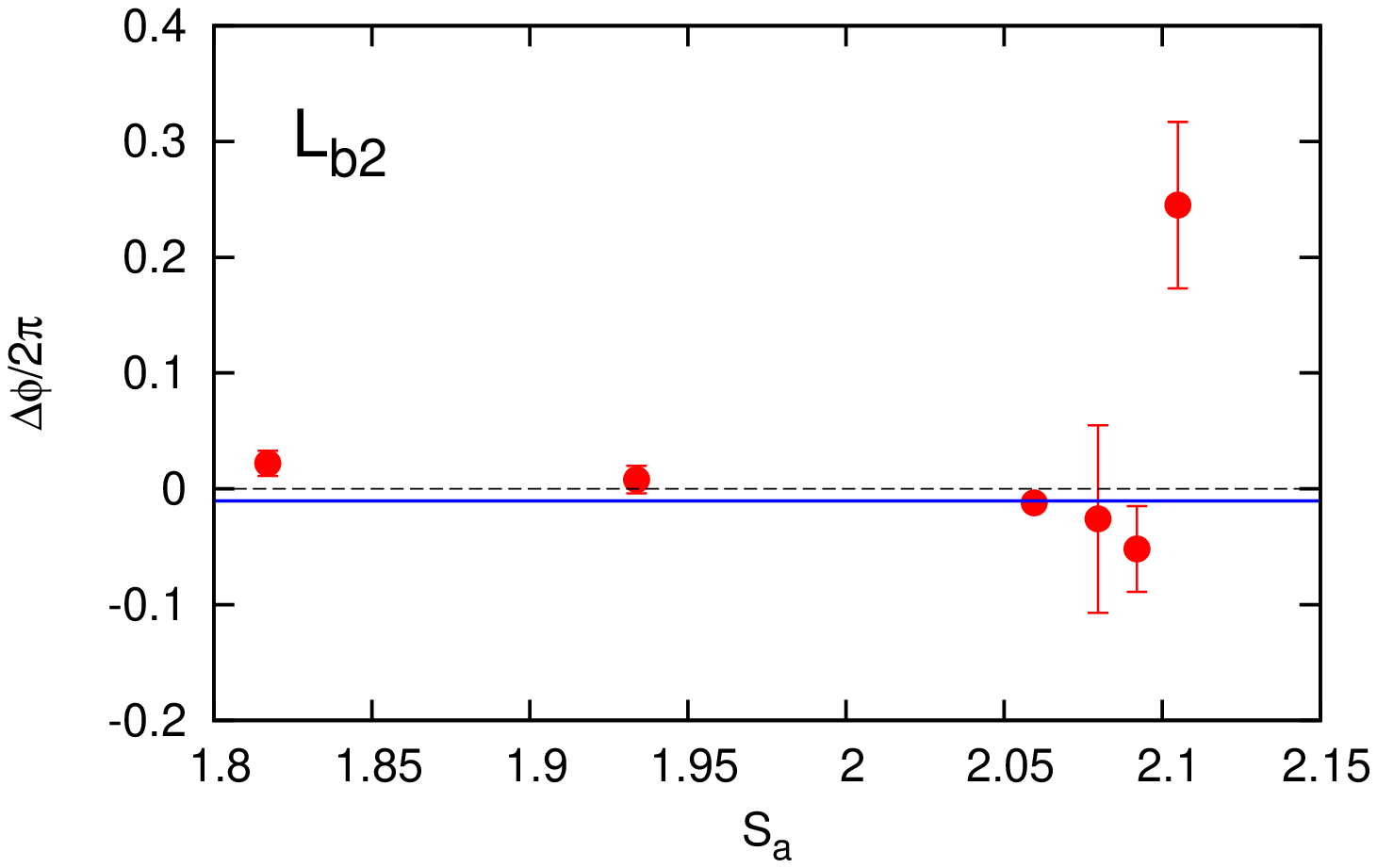}
}
{
    \includegraphics[width=0.325 \textwidth,angle=-90]{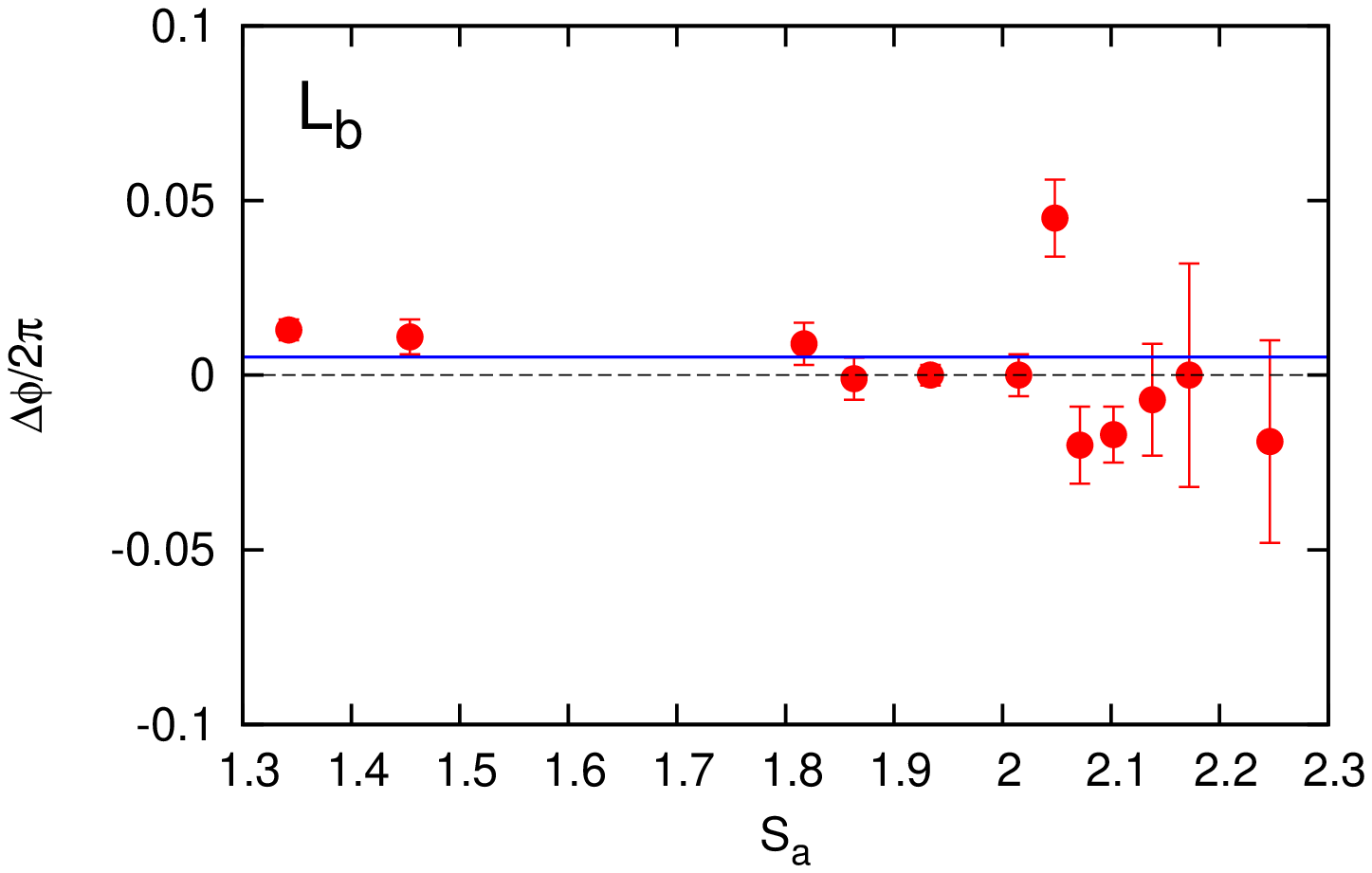}
}
{
    \includegraphics[width=0.325 \textwidth,angle=-90]{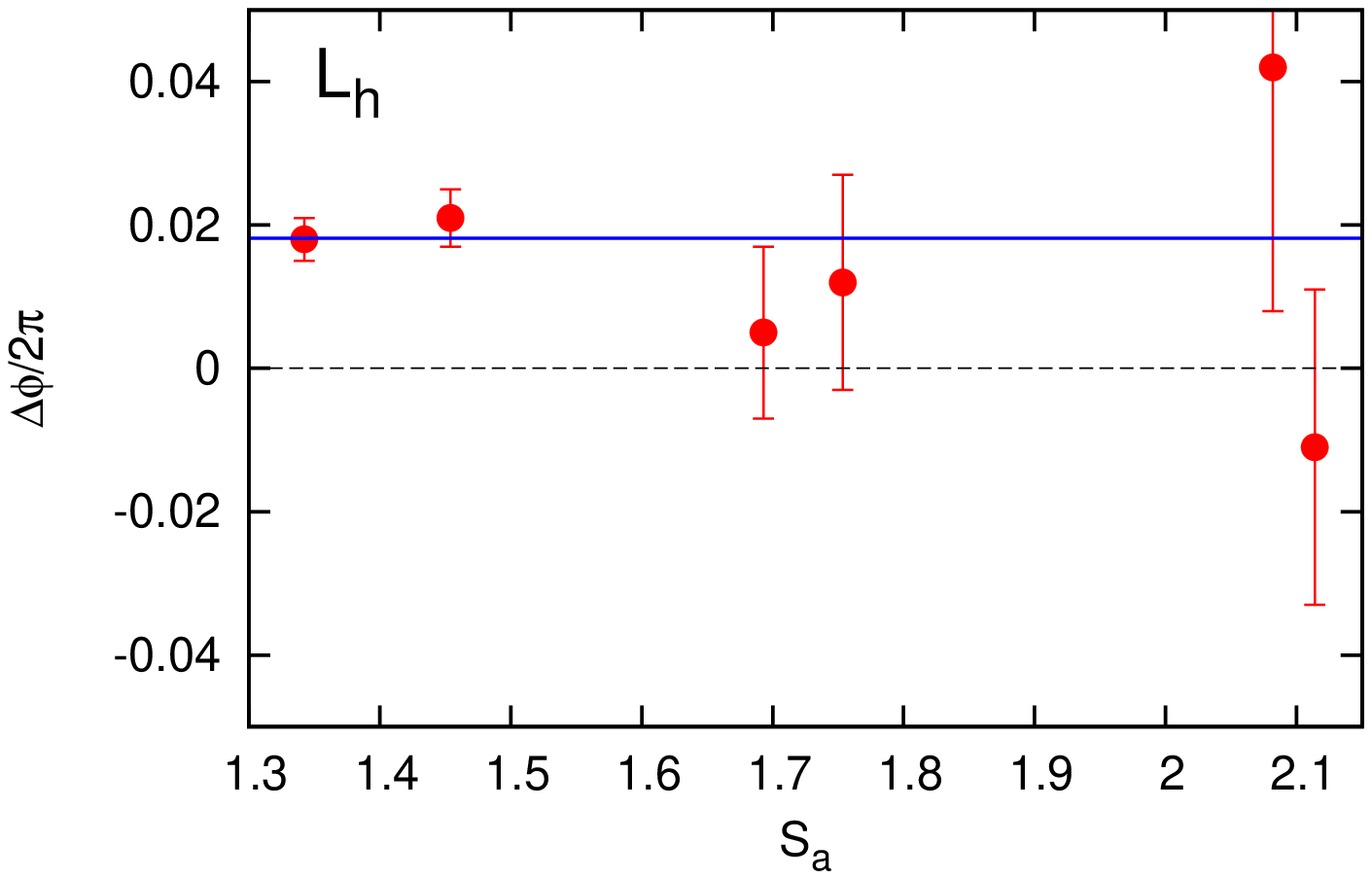}
}
{
    \includegraphics[width=0.325 \textwidth,angle=-90]{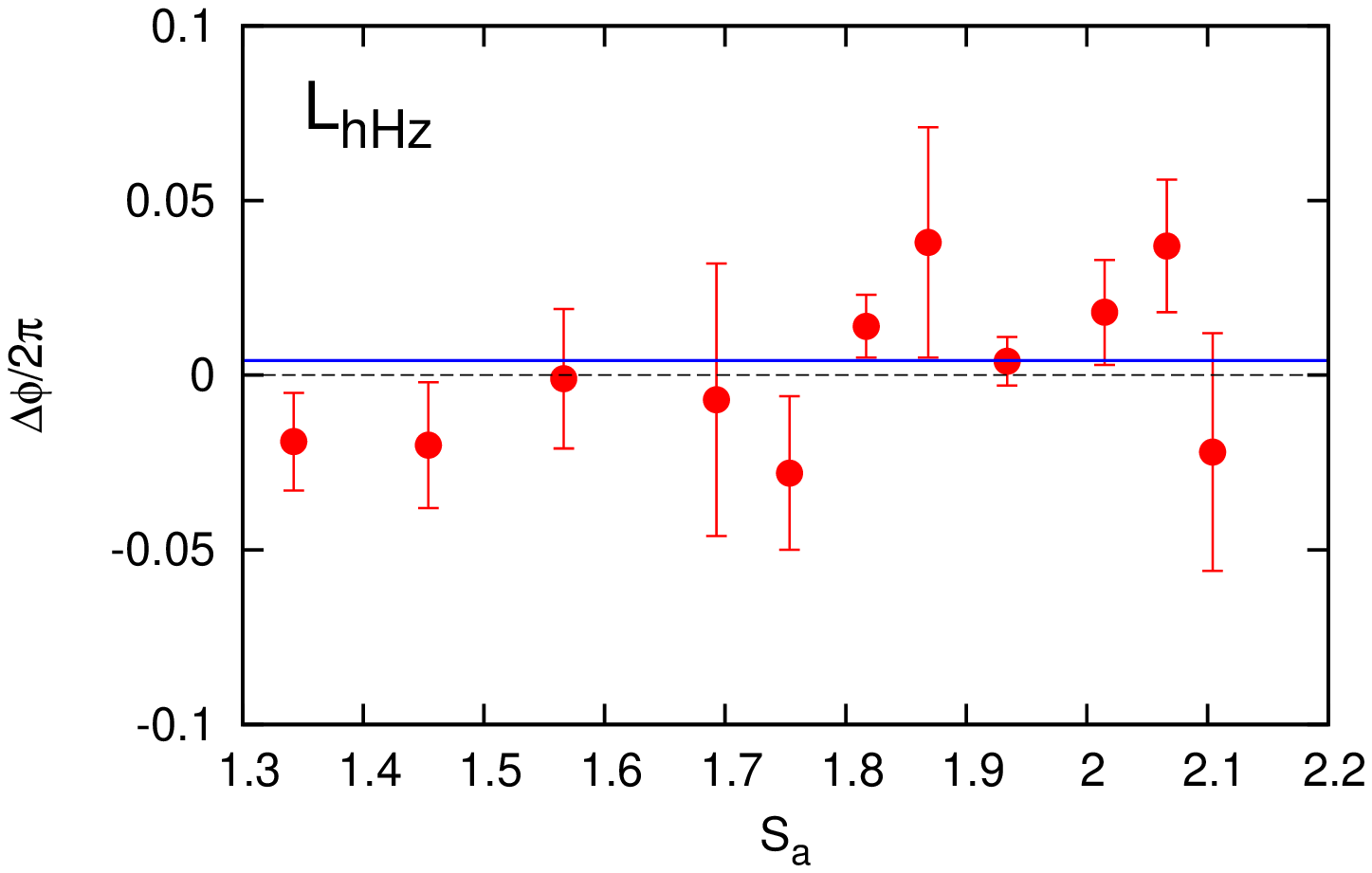}
}
{
    \includegraphics[width=0.325 \textwidth,angle=-90]{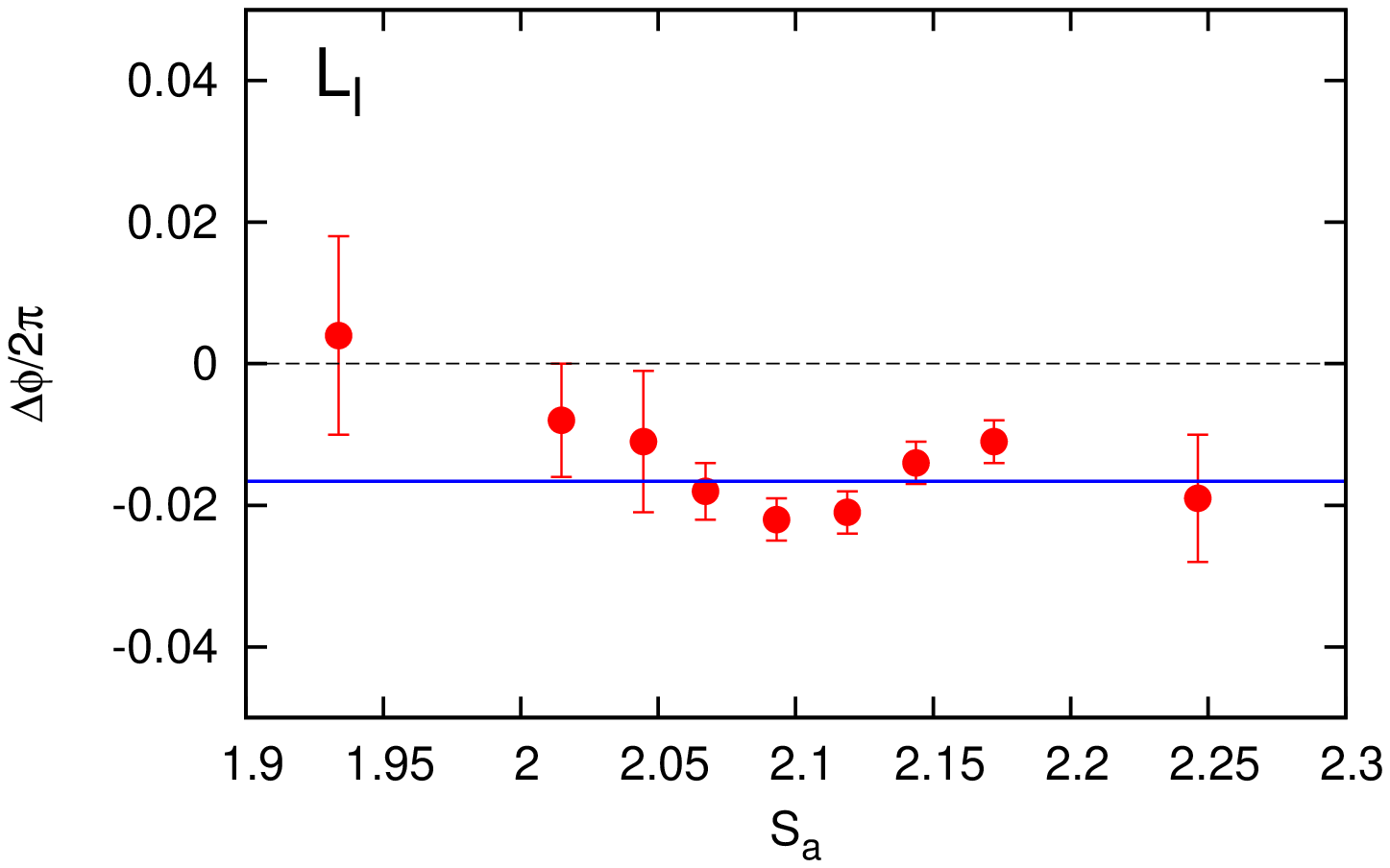}
}
{
    \includegraphics[width=0.325 \textwidth,angle=-90]{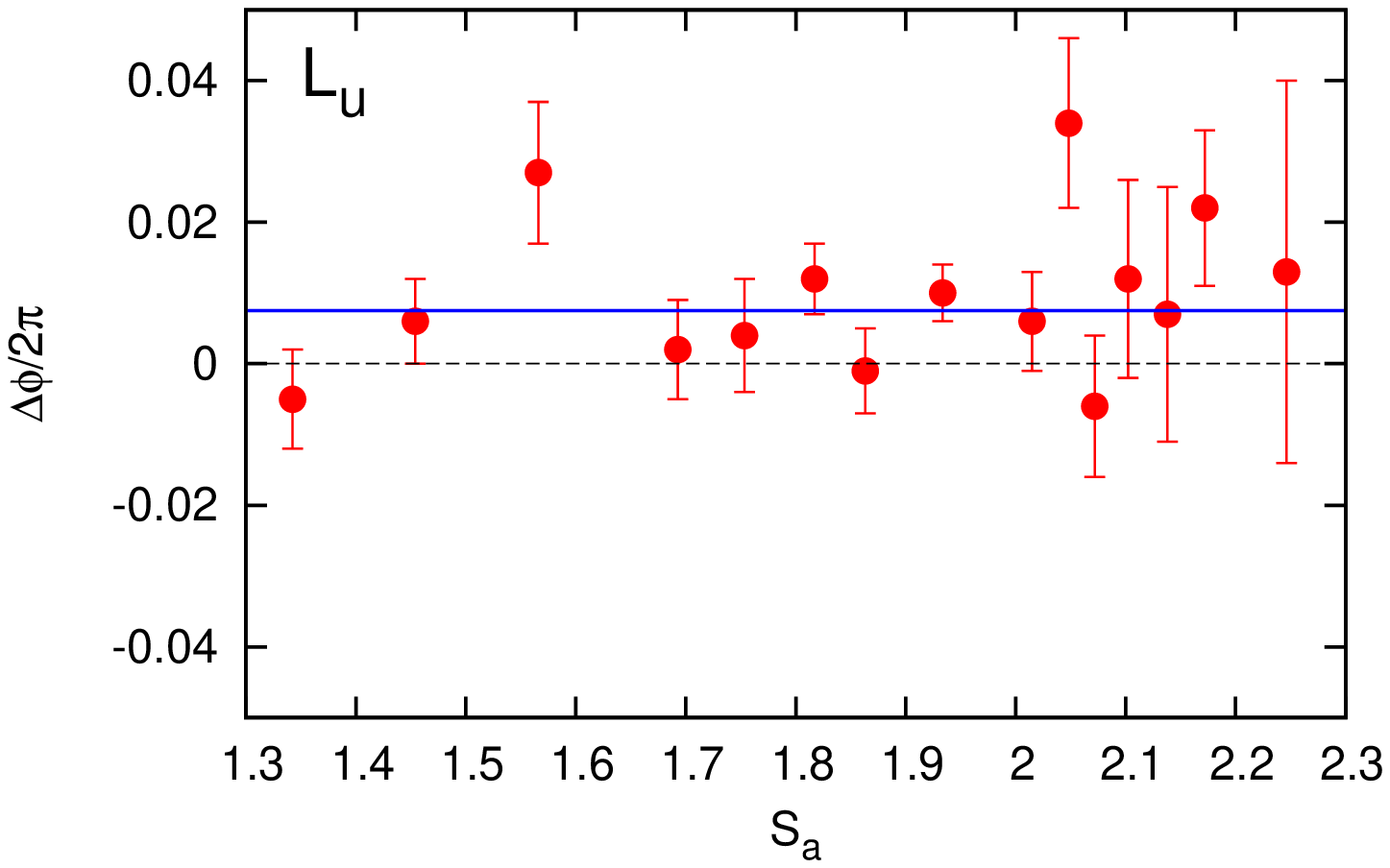}
}
\caption{Phase lags, in unit of $2\pi$, of the QPOs in Figure \ref{fig:phaseLagsVSfrequency} as a function of the parameter $S_{a}$ which is representative of the position of the source in the CCD. The blue lines are the best fit constant to the lags.}
\label{fig:phaseLagWithSa} 
\end{figure*}

In Figure \ref{fig:phaseLagWithSa} we plot the phase lags of the QPO specified in each of the panels as a function of the quantity $S_{a}$. 

%
%
%

We rebinned the data points in each panel of each figure individually in the following way: in Figure \ref{fig:phaseLagsVSfrequency}, we rebinned $L_{b2}$ in intervals of 0.8 Hz, $L_{b}$ in intervals of 4 Hz, $L_{h}$ in intervals of 4 Hz, $L_{hHz}$ in intervals of 10 Hz, $L_{l}$ in intervals of 34 Hz and $L_{u}$ in intervals of 35 Hz. This is approximately 10\% of the full range. In Figure \ref{fig:phaseLagWithSa}, we rebinned $L_{b2}$ in intervals of 0.025 in $S_{a}$, $L_{b}$ in intervals of 0.03 in $S_{a}$, $L_{h}$ in intervals of 0.05 in $S_{a}$, $L_{hHz}$ in intervals of 0.05 in $S_{a}$, $L_{l}$ in intervals of 0.025 in $S_{a}$ and $L_{u}$ in intervals of 0.03 in $S_{a}$. These values were chosen to avoid rebinning too much to lose information about any possible trend in the data. Note that two points close in frequency may be not close in $S_{a}$. Thus, the number of points in each panel of Figures \ref{fig:phaseLagsVSfrequency} and \ref{fig:phaseLagWithSa} may be different and small statistical differences are expected in some cases.

In Figures \ref{fig:phaseLagsVSfrequency} and \ref{fig:phaseLagWithSa} we show the phase lags {\it versus} frequency and {\it versus} $S_{a}$ and in Tables \ref{tab:statsLagsVSfreq} and \ref{tab:statsLagsVSsa} we show our statistical analysis. The main points are:

\begin{itemize}

\item $L_{b2}$ does not show any trend either with frequency or $S_{a}$. Although there is an outlier point at $\simeq 4.3$ Hz that suggests an increase of the phase lags from $-0.012$ at $0.63$ Hz to $0.245$ at $4.3$ Hz and then a decrease to $-0.048$ at $8.76$ Hz, this outlier is within $3\sigma$ from the other points. The same is true in relation to $S_{a}$. The average phase lag is negative.

\item $L_{b}$ does not show any trend either with frequency or $S_{a}$. The phase lags show larger fluctuations above 30 Hz than below 30 Hz and in relation to $S_{a}$ the phase lags vary more in the softest states. The average phase lag is positive.

\item $L_{h}$ does not show any trend either with frequency or $S_{a}$. The average phase lag is positive.

\item $L_{hHz}$ shows phase lags consistent in being constant and positive with frequency and $S_{a}$. Although it may appear that the phase lags are slightly positive for frequencies between 80 and 120 Hz and slightly negative for frequencies between 180 and 200 Hz (notice the opposite behaviour with $S_{a}$: slightly negative up to $S_{a}=1.8$ and positive above it), the F-tests probabilities in the two cases do not favour any trend.

\item The phase lags of $L_{l}$ are on average soft $-0.018\pm 0.001$. Although F-tests do not favour simple trends like linear or quadratic against the constant fit, this QPO seems to show a fluctuation pattern with both, frequency and $S_{a}$, that may be an indication of a more complex behaviour of the lags.

\item Finally, $L_{u}$ shows hard average phase lags ($0.007\pm 0.002$) and does not show any significant trend with frequency or $S_{a}$.

\end{itemize}

%
%
%
\begin{table*}
\caption{Statistics of Figure \ref{fig:phaseLagsVSfrequency}, phase lags {\it vs} frequency.}
\centering
\begin{tabular}{|c|c|c|c|c|c|c|}
\hline
\multicolumn{7}{|c|}{Frequency-dependent fits of the phase lags of 4U 1636--53}\\
\hline
~ & \multicolumn{2}{|c|}{const.} & \multicolumn{2}{|c|}{linear} & \multicolumn{2}{|c|}{quadratic} \\
\hline
~ & $\chi^{2}$ & DoF & $\chi^{2}$ & DoF & $\chi^{2}$ &  DoF \\
\hline
    $L_{b2}$ & 23.20392 & 4 & 14.47323 & 3 & 9.29348 & 2 \\
    $L_{b}$ & 40.78458 & 9 & 28.60968 & 8 & 28.59388 & 7 \\
    $L_{h}$ & 3.53457 & 3 & 3.19182 & 2 & 0.567889 & 1 \\
    $L_{hHz}$ & 11.57457 & 7 & 5.341212 & 6 & 5.03730 & 5 \\
    $L_{l}$ & 29.55393 & 9 & 29.43456 & 8 & 16.24865 & 7 \\
    $L_{u}$ & 18.68430 & 15 & 16.39764 & 14 & 16.35465 & 13 \\
\hline
\hline
\end{tabular}

\centering
\begin{tabular}{|c|c|c|c|c|c|c|}
\hline
\multicolumn{7}{|c|}{Frequency-dependent trends of the phase lags of 4U 1636--53}\\
\hline
~ & \multicolumn{2}{|c|}{const. vs linear} & \multicolumn{2}{|c|}{const. vs quadratic} & \multicolumn{2}{|c|}{linear vs quadratic} \\
\hline
~ & Fvalue & Prob & Fvalue & Prob & Fvalue &  Prob \\
\hline
    $L_{b2}$ & 1.80969 & 0.271195 & 1.4968 & 0.400513 & 1.11471 & 0.401765 \\
    $L_{b}$ & 3.40441 & 0.102242 & 1.49219 & 0.288549 & 0.00386796 & 0.952148 \\
    $L_{h}$ & 0.214768 & 0.688599 & 2.61203 & 0.400833 & 4.6205 & 0.277207 \\
    $L_{hHz}$ & 7.00218 & 0.0382241 & 3.24443 & 0.124949 & 0.301662 & 0.60647 \\
    $L_{l}$ & 0.0324435 & 0.861537 & 2.86599 & 0.123227 & 5.68056 & 0.048639 \\
    $L_{u}$ & 1.95231 & 0.184091 & 0.925897 & 0.420793 & 0.0341719 & 0.856194 \\
\hline
\hline
\end{tabular}
\label{tab:statsLagsVSfreq}
\end{table*}

%
%
%
\begin{table*}
\caption{Statistics of Figure \ref{fig:phaseLagWithSa}, phase lags {\it vs} $S_{a}$.}
\centering
\begin{tabular}{|c|c|c|c|c|c|c|}
\hline
\multicolumn{7}{|c|}{Frequency-dependent fits of the phase lags of 4U 1636--53}\\
\hline
~ & \multicolumn{2}{|c|}{const.} & \multicolumn{2}{|c|}{linear} & \multicolumn{2}{|c|}{quadratic} \\
\hline
~ & $\chi^{2}$ & DoF & $\chi^{2}$ & DoF & $\chi^{2}$ &  DoF \\
\hline
    $L_{b2}$ & 25.55025 & 5 & 14.34936 & 4 & 14.31816 & 3 \\
    $L_{b}$ & 40.67437 & 11 & 25.34820 & 10 & 24.78186 & 9 \\
    $L_{h}$ & 4.126740 & 5 & 3.164756 & 4 & 2.839491 & 3 \\
    $L_{hHz}$ & 13.50180 & 10 & 7.468119 & 9 & 7.441792 & 8 \\
    $L_{l}$ & 13.45464 & 8 & 13.15979 & 7 & 8.74152 & 6 \\
    $L_{u}$ & 19.69842 & 14 & 17.73109 & 13 & 17.700 & 12 \\
\hline
\hline
\end{tabular}

\centering
\begin{tabular}{|c|c|c|c|c|c|c|}
\hline
\multicolumn{7}{|c|}{Frequency-dependent trends of the phase lags of 4U 1636--53}\\
\hline
~ & \multicolumn{2}{|c|}{const. vs linear} & \multicolumn{2}{|c|}{const. vs quadratic} & \multicolumn{2}{|c|}{linear vs quadratic} \\
\hline
~ & Fvalue & Prob & Fvalue & Prob & Fvalue &  Prob \\
\hline
    $L_{b2}$ & 3.12234 & 0.151968 & 1.1767 & 0.419506 & 0.00653715 & 0.940651 \\
    $L_{b}$ & 6.04626 & 0.0337391 & 2.88583 & 0.107562 & 0.205677 & 0.660914 \\
    $L_{h}$ & 1.21587 & 0.332053 & 0.680007 & 0.570755 & 0.343651 & 0.598917 \\
    $L_{hHz}$ & 7.27133 & 0.0245323 & 3.25728 & 0.0922876 & 0.0283018 & 0.870577 \\
    $L_{l}$ & 0.156838 & 0.703879 & 1.61749 & 0.274249 & 3.03261 & 0.132247 \\
    $L_{u}$ & 1.4424 & 0.251177 & 0.677431 & 0.526322 & 0.021078 & 0.886977 \\
\hline
\hline
\end{tabular}
\label{tab:statsLagsVSsa}
\end{table*}

We found approximately the same patterns of the phase lags shown as a function of the frequency of the kHz QPOs as a function of the $S_{a}$, as shown in the bottom panels of Figure \ref{fig:phaseLagWithSa}. The phase lags of $L_{l}$ decrease from $\simeq 0.004$ at $S_{a}$ $\simeq 1.9$ to $\simeq -0.02$ at $S_{a}$ $\simeq 2.12$ and then they remain constant or increase slightly. The phase lags of $L_{u}$, in turn, fluctuate around the average value as $S_{a}$ increases. These results were already pointed out in \cite{deAvellar01}.

The region of the CCD in which $2.05\leq S_{a} \leq 2.15$ marks the transition point to the ``softest states'' (from $\sim 0.03~L_{edd}$ to $\sim 0.3~L_{edd}$ and from 1--100 Hz fractional rms $\sim 10\%$ to $\sim 1\%$, for example, at constant HC, estimated from \citealt{linaresThesis}). We also detect both kHz QPOs in this region. See \cite{linaresThesis} for an extensive discussion on the relations between luminosity, spectra and variability. In this region, while the phase lags of $L_{u}$ remains fairly constant and hard, the phase lags of $L_{l}$ reach the softest values, forming a valley. 

The results for the kHz QPOs and for $L_{b}$ and $L_{hHz}$ obtained here are consistent with those obtained, respectively, by \cite{deAvellar01} and by \cite{deAvellar02}, the latter using a smaller dataset, for which we selected the data on the basis of the frequency of the kHz QPOs as in \cite{deAvellar01}. The selection we used in this work, based on the position of the source on the CCD, intrinsically produces more dispersion than the one used in the previous works.

In Table \ref{tab:constantPhaseLagsValues} we show constant fits of the phase lags as a function of $S_{a}$ of all QPOs in order to compare them with each other. We see from Table \ref{tab:constantPhaseLagsValues} that the average phase lags are soft in the cases of $L_{b2}$ and $L_{l}$. The phase lags of $L_{b2}$ are significantly different from the phase lags of $L_{b}$, $L_{h}$ and $L_{u}$, the phase lags of $L_{b}$ are significantly different from the lags of $L_{h}$ and $L_{l}$, the phase lags of $L_{h}$ are significantly different from the lags of $L_{l}$, and the phase lags of $L_{l}$ are significantly different from the lags of $L_{u}$. The phase lags of $L_{b}$, $L_{hHz}$ and $L_{u}$ are consistent with each other within $3\sigma$.

%
%
%
\begin{table}
\caption{Average phase lags and time lags as function of $S_{a}$ (see Figure \ref{fig:phaseLagWithSa}) for the $L_{b2}$, $L_{b}$, $L_{h}$, $L_{hHz}$, $L_{l}$ and $L_{u}$ QPOs of the NS-LMXB 4U 1636--53 of all photons in the broad band 12-20 keV relatively to all photons in the broad band 4-12 keV.}
\centering
\begin{tabular}{|c|c|c|c|c}
\hline
\multicolumn{5}{|c|}{Average lags of 4U 1636--53}\\
\hline
QPO & phase lag [$2\pi$ rad] & $\chi^{2}$ & dof & time lag [msec] \\
\hline
    $L_{b2}$ & $-0.010\pm 0.002$ & 25.6 & 5 & $-8.7\pm2.1$ \\
    $L_{b}$ & $0.005\pm 0.002$ & 40.7 & 11 & $0.05\pm0.09$ \\
    $L_{h}$ & $0.018\pm 0.002$ & 4.1 & 5 & $0.8\pm0.1$ \\
    $L_{hHz}$ & $0.004\pm 0.004$ & 13.5 & 10 & $0.008\pm0.035$ \\
    $L_{l}$ & $-0.017\pm 0.001$ & 13.5 & 8 & $-0.020\pm0.002$ \\
    $L_{u}$ & $0.008\pm 0.002$ & 19.7 & 14 & $0.010\pm0.002$ \\
\hline
\hline
\end{tabular}
\label{tab:constantPhaseLagsValues}
\end{table}

In Table \ref{tab:constantPhaseLagsValues} we also give the average {\it time} lags as a function of $S_{a}$ for each QPO, which we will use later. Except for the $L_{b2}$ ($-8.7$ msec) and $L_{h}$ ($0.8$ msec), we see from the Table that the values are comparable for all QPOs and only in the cases of $L_{b2}$ and $L_{l}$ we see soft average lags.

A point should be mentioned now. The phase lags and their errors (and the corresponding time lags) were calculated following the prescriptions in \cite{vaughan01} and \cite{kaaret01}. In Table \ref{tab:constantPhaseLagsValues} we show the weighted average of the phase and time lags for each QPO and their respective error bars.

The {\it time} lags of $L_{b2}$ are inconsistent with the time lags of all other QPOs by more than $3\sigma$. The time lags of $L_{b}$ are consistent within $3\sigma$ with the time lags of all other QPOs but $L_{h}$. The time lags of $L_{h}$, in turn, are inconsistent within $3\sigma$ with the time lags of $L_{hHz}$, $L_{l}$ and $L_{u}$.  At last, the time lags of $L_{l}$ are different from the time lags of $L_{u}$ by more than $3\sigma$, a result previously known \citep{deAvellar01}.

\subsection{Energy dependence of the phase lags}
\label{res2}

\begin{figure}
\centering
{
    \includegraphics[width=0.48 \textwidth,angle=-90]{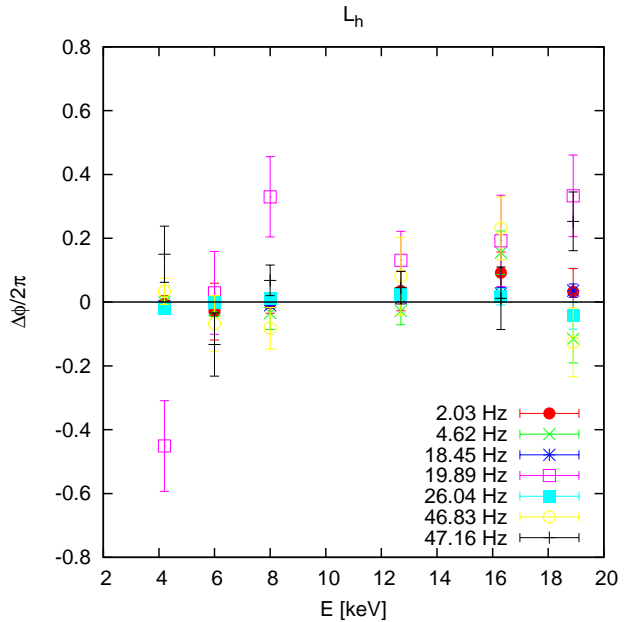}
}
\caption{Phase lags {\it vs} energy for the $L_{h}$ QPO of 4U 1636--53 as its frequency drifts along the CCD. Notice that the lags are all consistent each other within the error bars since the phase lags do not depend upon the frequency. Therefore we could average the phase lags in order to get the plots shown in Figure \ref{fig:phaseLagsVSenergy}.}
\label{fig:lags-vs-E-all-Lh}  
\end{figure}

The results in Section \ref{res1} (using only two broad energy bands) show that the phase lags do not depend strongly on the frequency of the QPO or $S_{a}$, except maybe for the lower kHz QPO (bottom left panel of Figure \ref{fig:phaseLagsVSfrequency} and \ref{fig:phaseLagWithSa}, respectively). Based on this, we averaged the phase lags over frequency to calculate the phase lags versus energy for each QPO under consideration in the following way: First we calculated the phase lags of each QPO in each box for the energy bands previously defined relative to 10.2 keV. This procedure produced the energy dependence of the phase lags of each QPO present in each of the 37 boxes. We then averaged these ``energy dependencies'': for the same energy band, we weight-averaged the phase lags of a given QPO of the box where it appears. Because the phase lags do not depend strongly on the frequency, the frequency drift of the QPOs over the CCD does not have a big influence in this averaging: the lags {\it versus} energy for each frequency of a given QPO are consistent within the errors (see Figure \ref{fig:lags-vs-E-all-Lh}). Notice that, since the phase lags of $L_{l}$ are not independent of frequency, this procedure is not completely correct for this QPO.

%
%
%

\begin{figure*}
\centering
{
    \includegraphics[width=0.325 \textwidth,angle=-90]{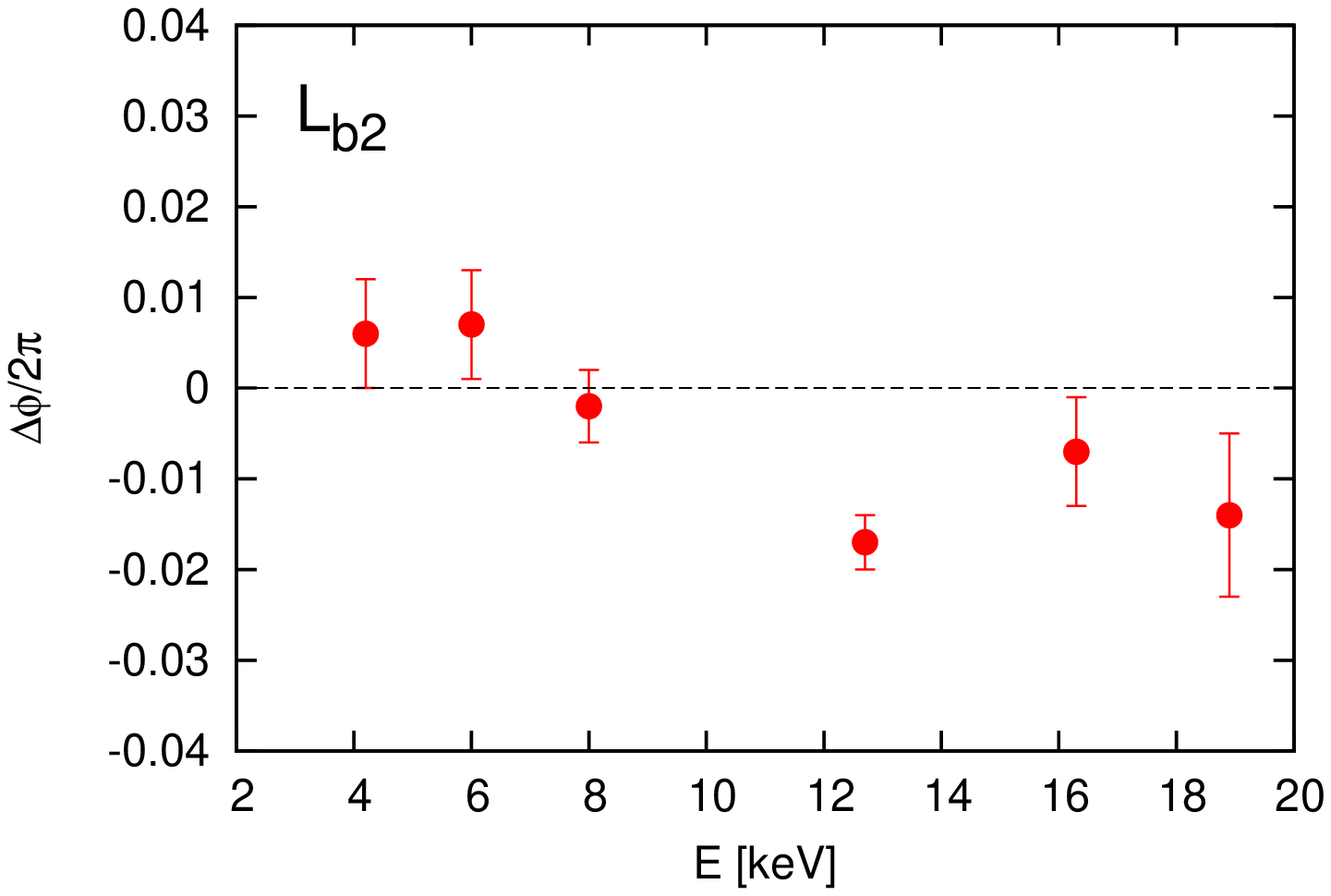}
}
{
    \includegraphics[width=0.325 \textwidth,angle=-90]{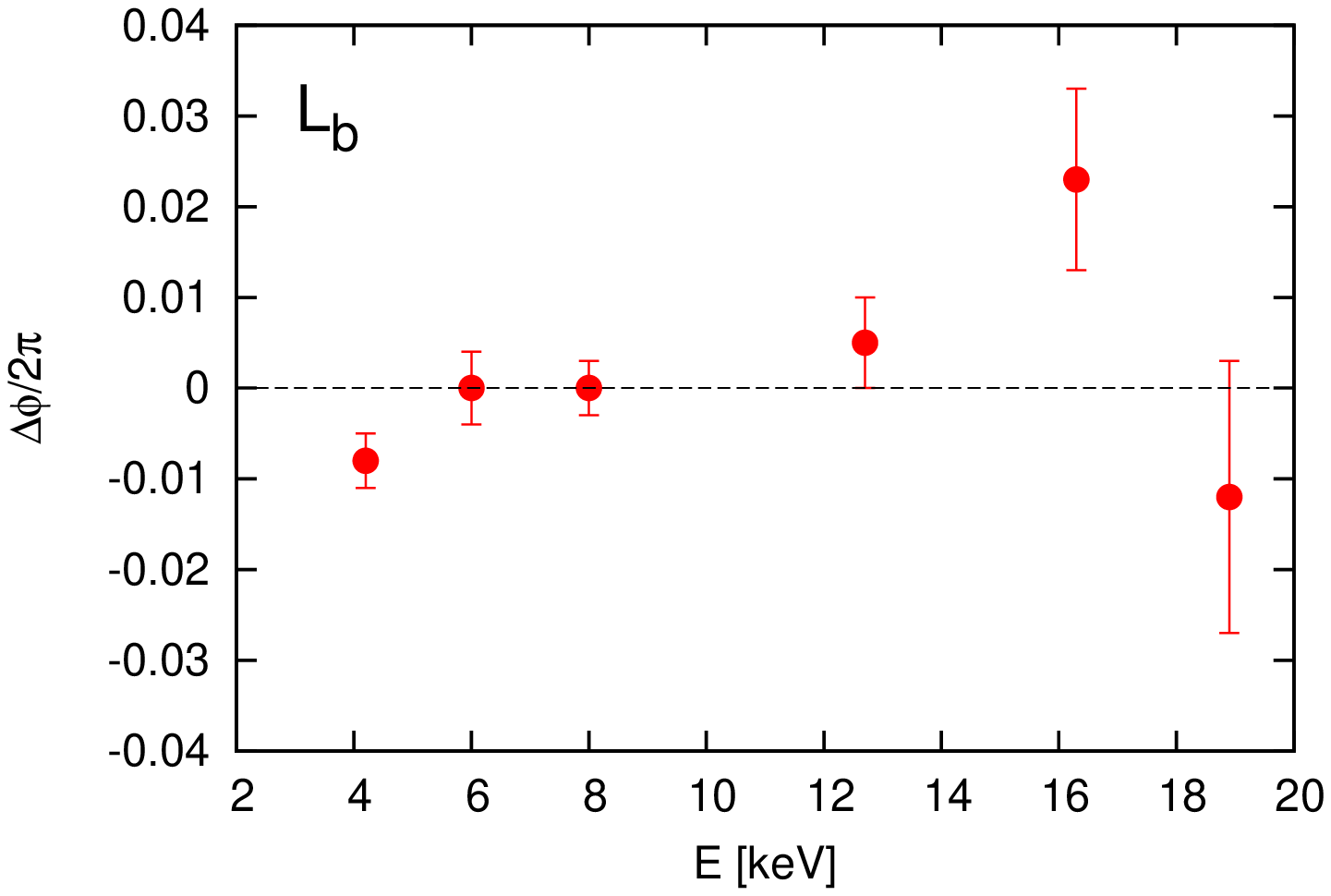}
}
{
    \includegraphics[width=0.325 \textwidth,angle=-90]{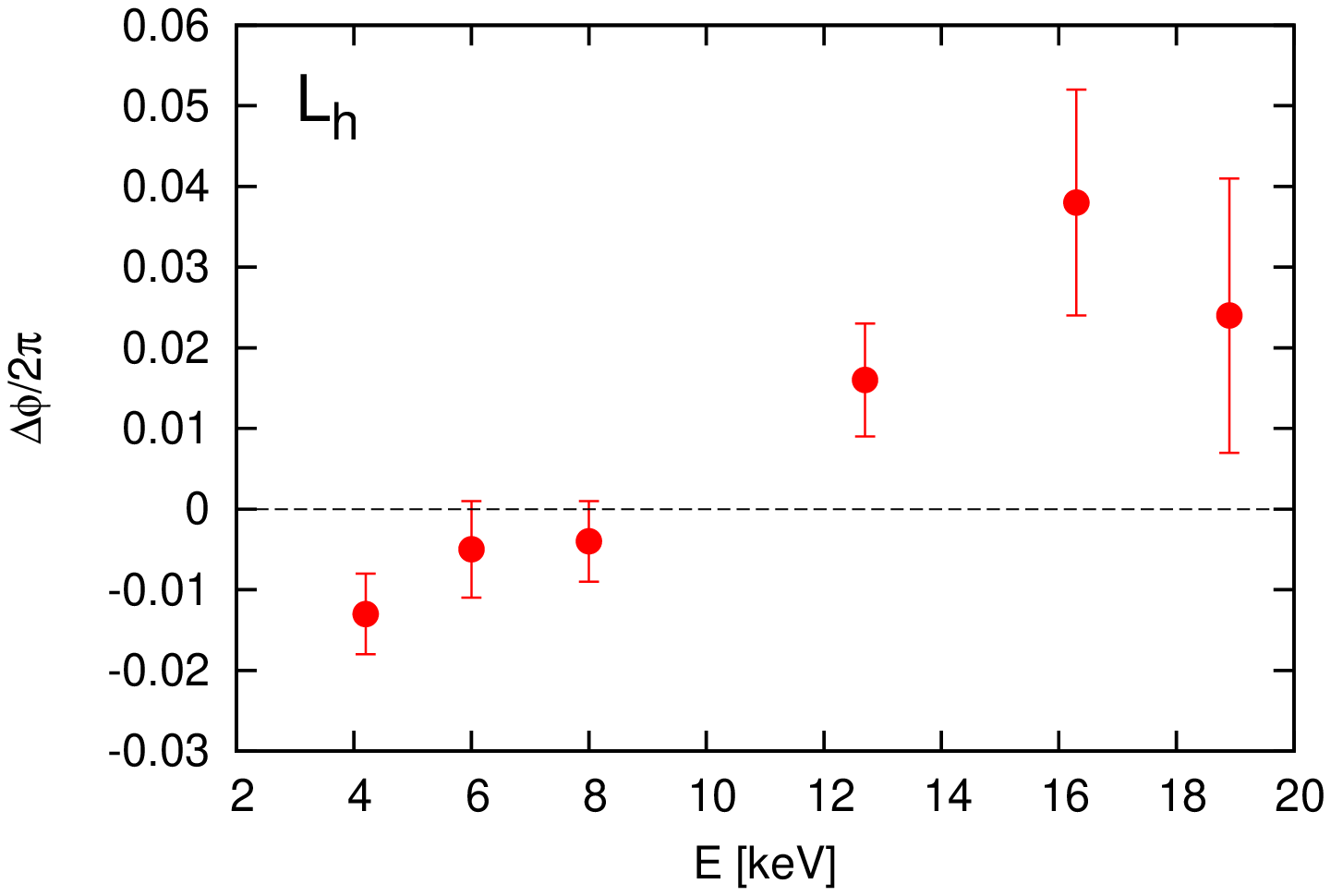}
}
{
    \includegraphics[width=0.325 \textwidth,angle=-90]{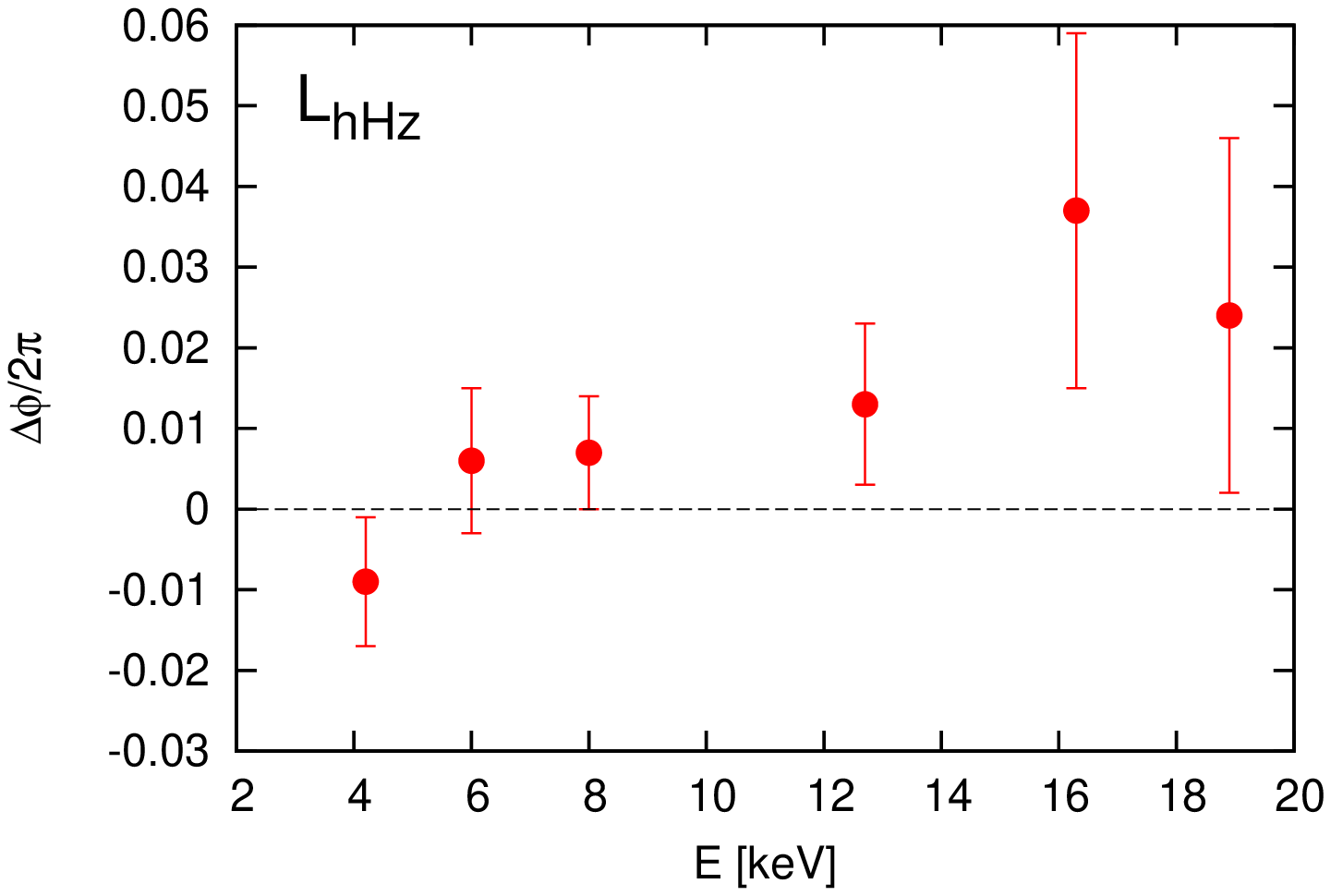}
}
{
    \includegraphics[width=0.325 \textwidth,angle=-90]{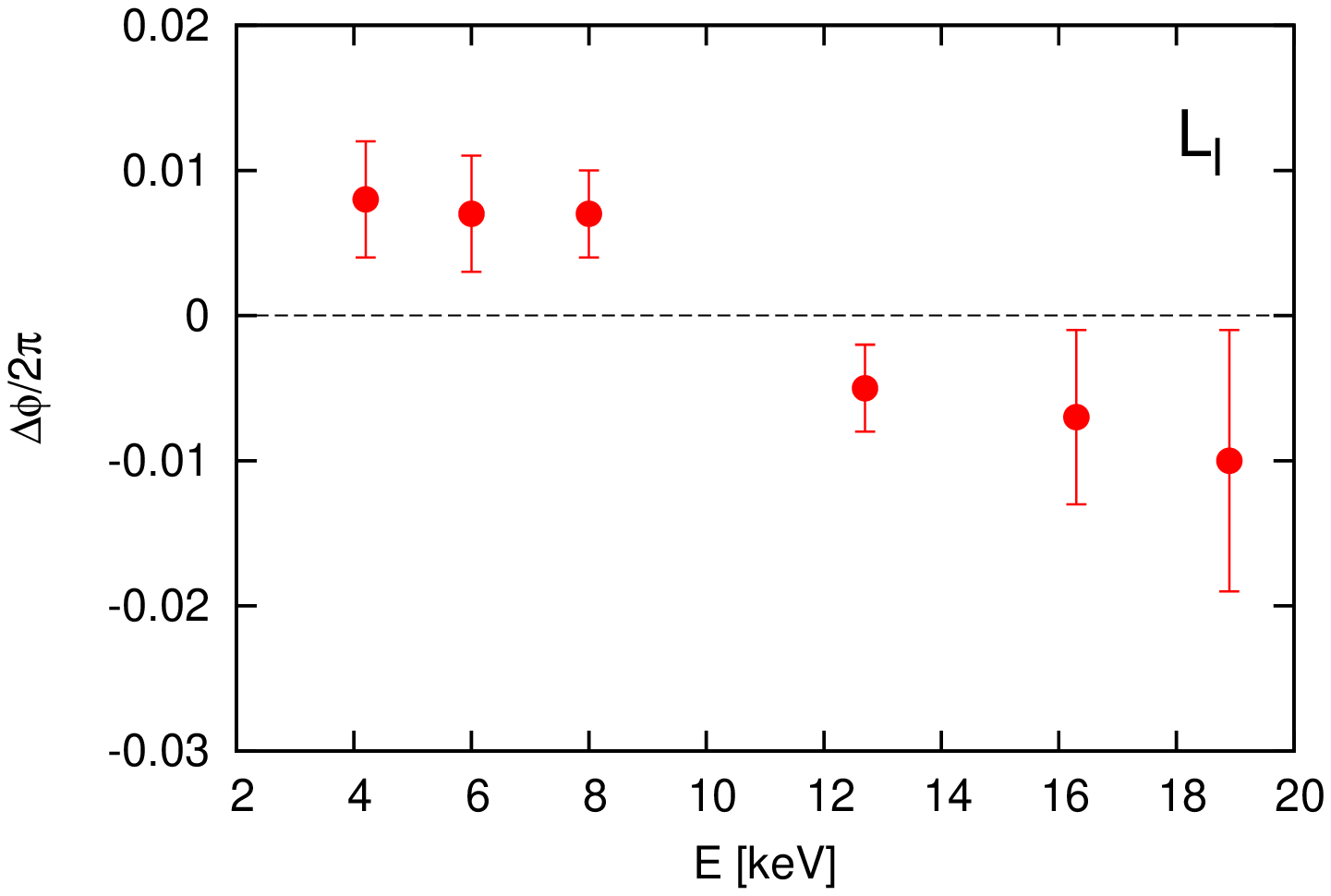}
}
{
    \includegraphics[width=0.325 \textwidth,angle=-90]{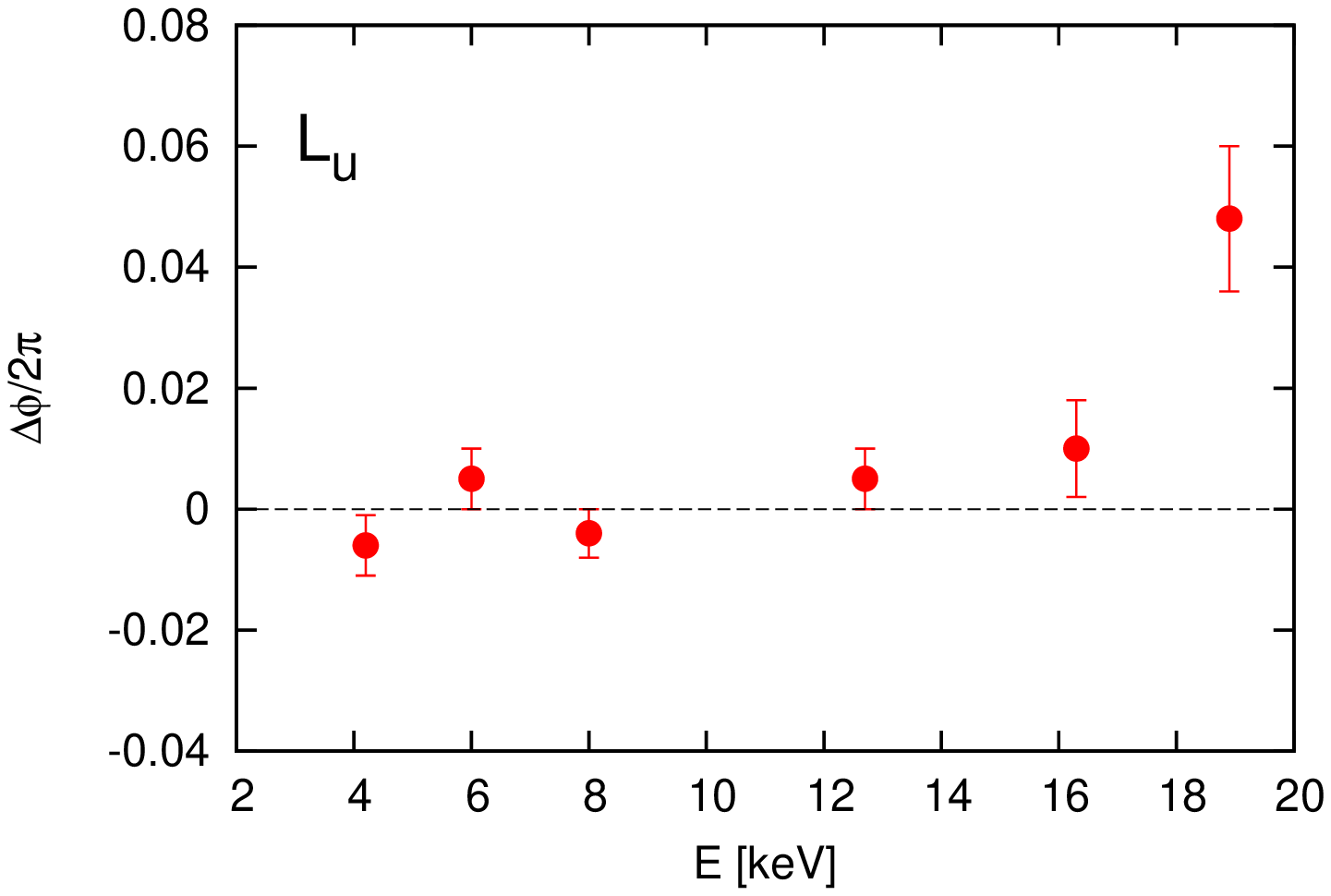}
}
\caption{Phase lags, in unit of $2\pi$, of the different QPO components of 4U 1636--53 as a function of energy. The lags represent the delay of photons in the bands whose average energies are given by the values in the $x$-axis relative to the photons in the band whose average energy is 10.2 keV. The panels show the same QPOs as in Figure \ref{fig:phaseLagsVSfrequency}.}
\label{fig:phaseLagsVSenergy}  
\end{figure*}

The results of this procedure are shown in Figure \ref{fig:phaseLagsVSenergy}. In Table \ref{tab:statsLagsVSenergy} we show our statistical analysis. The main points are:

\begin{itemize}

\item $L_{b2}$: F-tests tell us that a linear or quadratic function are not better than the constant fit. However, there is an abrupt decrease of the lags at $12.7$ keV, more than $3\sigma$ below the constant fitted value ($-0.017 \pm 0.003$) that seems well constrained.

\item $L_{b}$: There is no significant trend with energy for this QPO.

\item $L_{h}$ shows one of the strongest energy dependence we have found in this study with an F-test probability for a constant versus a linear fit of $0.002$. The rate of increase of the phase lags is $(0.0030 \pm 0.0007)/2\pi$ per keV.

\item $L_{hHz}$: Formally, the phase lags for this QPO show no trend with energy (the F-test for a linear against a constant fit is $\simeq 0.02$).

\item The phase lags of $L_{l}$ show a marginally decreasing trend with energy with F-test value for a constant versus linear of 0.004. Formally this value is a bit larger than the 3$\sigma$ value and is probably due to the larger error bars on the data here. This trend is already known from previous works.

\item Finally, the phase lags of the upper kHz QPO, $L_{u}$, are constant with energy with F-test for a constant against a linear of 0.11.

\end{itemize}

%
%
%

%
%
%
\begin{table*}
\caption{Statistics of Figure \ref{fig:phaseLagsVSenergy}, phase lags {\it vs} energy.}
\centering
\begin{tabular}{|c|c|c|c|c|c|c|}
\hline
\multicolumn{7}{|c|}{Energy-dependent trends of the phase lags of 4U 1636--53}\\
\hline
~ & \multicolumn{2}{|c|}{const.} & \multicolumn{2}{|c|}{linear} & \multicolumn{2}{|c|}{quadratic} \\
\hline
~ & $\chi^{2}$ & DoF & $\chi^{2}$ & DoF & $\chi^{2}$ &  DoF \\
\hline
    $L_{b2}$ & 23.34975 & 5 & 8.94780 & 4 & 4.12626 & 3 \\
    $L_{b}$ & 13.26755 & 5 & 5.87532 & 4 & 4.68939 & 3 \\
    $L_{h}$ & 22.33995 & 5 & 1.704112 & 4 & 1.697379 & 3 \\
    $L_{hHz}$ & 6.65180 & 5 & 1.379504 & 4 & 1.166424 & 3 \\
    $L_{l}$ & 16.05085 & 5 & 1.712796 & 4 & 1.698921 & 3 \\
    $L_{u}$ & 21.22390 & 5 & 10.38140 & 4 & 6.31083 & 3 \\
\hline
\hline
\end{tabular}

\centering
\begin{tabular}{|c|c|c|c|c|c|c|}
\hline
\multicolumn{7}{|c|}{Energy-dependent trends of the phase lags of 4U 1636--53}\\
\hline
~ & \multicolumn{2}{|c|}{const. vs linear} & \multicolumn{2}{|c|}{const. vs quadratic} & \multicolumn{2}{|c|}{linear vs quadratic} \\
\hline
~ & Fvalue & Prob & Fvalue & Prob & Fvalue &  Prob \\
\hline
    $L_{b2}$ & 6.43821 & 0.0641604 & 6.98823 & 0.0742868 & 3.5055 & 0.157894 \\
    $L_{b}$ & 5.03273 & 0.0882909 & 2.74392 & 0.210129 & 0.758705 & 0.447833 \\
    $L_{h}$ & 48.4378 & 0.00224003 & 18.2422 & 0.0209433 & 0.0119001 & 0.92002 \\
    $L_{hHz}$ & 15.2875 & 0.0173949 & 7.05409 & 0.0734304 & 0.548034 & 0.512795 \\
    $L_{l}$ & 33.4846 & 0.0044319 & 12.6715 & 0.034436 & 0.0245008 & 0.885558 \\
    $L_{u}$ & 4.17766 & 0.110449 & 3.54464 & 0.162141 & 1.93504 & 0.258424 \\
\hline
\hline
\end{tabular}
\label{tab:statsLagsVSenergy}
\end{table*}

As with the frequency dependence, the results for the energy dependence of the phase lags of the kHz QPOs are consistent with the ones obtained in \cite{deAvellar01} using a different way to select the data \cite[see Figure 3 in][]{deAvellar01}.

%
%
%

It is interesting to note, based on the shapes of the lag vs E plots, that the trend with energy, being marginal or not, is an increasing trend for $L_{b}$, $L_{h}$, $L_{hHz}$, and $L_{u}$, but is decreasing for $L_{l}$ and possibly for $L_{b2}$. In the case of $L_{b2}$ the question arises if the phase lags of $L_{b2}$ may be linearly correlated with the phase lags of $L_{l}$, although the F-test probability of 0.026 indicates the contrary. The same question arises about a possible anti-correlation between the phase lags of $L_{h}$ and $L_{l}$ (F-test probability 0.005).

We therefore compared the slopes of the lag vs E linear fits for each QPO, and found that we can separate the components into two groups having slopes consistent with each other: On the one hand $L_{b2}$ and $L_{l}$, and on the other hand $L_{b}$, $L_{h}$, $L_{hHz}$, and $L_{u}$. (In the comparison we are performing now it is not important if we can or can not distinguish the linear trend of the constant with the F-tests as discussed above.) The slopes are shown in Table \ref{tab:slopes}.

\begin{table}
\caption{Slopes of the linear fits to the phase lags {\it vs} energy in Figure \ref{fig:phaseLagsVSenergy}.}
\centering
\begin{tabular}{|c|c|}
\hline
QPO & Slope   \\
\hline
    $L_{b2}$ & -0.0018 $\pm$ 0.0005 \\
    $L_{b}$ & 0.0014 $\pm$ 0.0005  \\
    $L_{h}$ & 0.0033 $\pm$ 0.0007  \\
    $L_{hHz}$ & 0.0024 $\pm$ 0.0011 \\
    $L_{l}$ & -0.0015 $\pm$ 0.0004  \\
    $L_{u}$ & 0.0018 $\pm$ 0.0005 \\
\hline
\hline
\end{tabular}
\label{tab:slopes}
\end{table}

In order to further investigate the slopes we used a totally different statistics. We calculated the Pearson Product Moment Correlation (PPMC) coefficient\footnote{This coefficient varies between -1 and 1.}, $r$, for all 15 possible combinations of lag vs lag. This correlation coefficient is a statistic that calculates the actual relationship between two variables. The values of $r$ are show in the Table \ref{tab:rValues}

\begin{table}
\caption{Pearson Product Moment Correlation coefficient for the energy dependence phase lag {\it vs} phase lag for all combinations of QPOs. In order to show positive correlation, $r\geq 0.955$ since we do not take as significant anything less than 3$\sigma$, which is a probability of 0.003 (or 99.7\%) for a Gaussian distribution. Notice that the only correlation we found was $L_{h}$/$L_{hHz}$ with $r=0.965$.}
\centering
\begin{tabular}{|c|c|}
\hline
QPO & PPMC coefficient ($r$)   \\
\hline
    $L_{b2}$/$L_{b}$ & -0.147  \\
    $L_{b2}$/$L_{h}$ & -0.733  \\
    $L_{b2}$/$L_{hHz}$ & -0.616  \\
    $L_{b2}$/$L_{l}$ & 0.870  \\
    $L_{b2}$/$L_{u}$ & -0.566 \\
    $L_{b}$/$L_{h}$ & 0.573  \\
    $L_{b}$/$L_{hHz}$ & 0.611  \\
    $L_{b}$/$L_{l}$ & -0.259  \\
    $L_{b}$/$L_{u}$ & -0.290  \\
    $L_{h}$/$L_{hHz}$ & 0.965  \\
    $L_{h}$/$L_{l}$ & -0.927 \\
    $L_{h}$/$L_{u}$ & 0.583  \\
    $L_{hHz}$/$L_{l}$ & -0.836  \\
    $L_{hHz}$/$L_{u}$ & 0.566 \\
    $L_{l}$/$L_{u}$ & -0.762  \\
\hline
\hline
\end{tabular}
\label{tab:rValues}
\end{table}

We have not found any correlation between the lags except in the case of the phase lags of $L_{h}$ in relation to the phase lags of $L_{hHz}$. Even so, we still can separate the lags in two groups as mentioned above.

\section{Discussion}
\label{dis}

%
%

We analysed, for the first time, the frequency and energy dependence of the phase lags of all the QPOs in 4U 1636--53 as a function of the spectral state of the source. For this we used the 511 observations of this source with the Rossi X-ray Timing Explorer in which \cite{sanna02} detected kHz QPOs. Except for the lower kHz QPO, the phase lags of all the other QPOs are independent of the frequency. Except for the lower kHz QPO and the hump QPO, the phase lags of all the other QPOs are independent of the energy.

\subsection{Frequency dependence}
\label{disFreq}


We see from Figure \ref{fig:phaseLagWithSa} that the phase lags are practically independent of $S_{a}$, except for $L_{l}$ (for which we see a complex behaviour) and may be for $L_{hHz}$. It has been suggested \citep{esin01,meyerHofmeister01,done01} that the accretion flow may have two components, $\dot{M_{d}}$ and $\dot{M_{c}}$, the accretion rate of the disc and of the corona, respectively. In a scenario where hard lags are produced by a Comptonized medium and soft lags are produced by reflection off a cold disc \citep{falanga01}, a possible explanation in the case of the lower kHz QPO lags is that if $\dot{M_{d}}$ increases while $\dot{M_{c}}$ decreases in a transition from hard states to soft states\footnote{Notice that this transition represents the source going from box 27 to box 20 and below, from where we begin to see the lower kHz QPO with its soft lags.}, the disc (thermal) component would partially suppress the non-thermal (corona) component. This partial suppression of the Comptonized medium would have as consequence the dominance of the soft lags producer component. Maybe the inner flow configuration that sets all these transitions is also responsible for these fluctuations of the phase lags of $L_{l}$ to softest values. This, of course, would not prevent the hard lags of the upper which would ``keep its constancy'' responding through compensations in the properties of the coronal component (see, e.g., expression \ref{eq1}).

%
%

%
%

In the context of models for the lags that involve reflection off the disc or Comptonization, light travel time arguments ($c\Delta t$, related to the time lags) are useful to give an upper limit to the size of the medium in which the time lags are produced. In Table \ref{tab:constantPhaseLagsValues} we give the average values of the time lags of each QPO. From those values we can roughly estimate the scale size, {\it a}, of the medium where the lags of each QPO are produced,

\begin{equation}
\label{eq1}
a\sim c\Delta t\frac{k_{b}T_{e}}{m_{e}c^{2}}\frac{4\tau}{ln(E_{2}/E_{1})}.
\end{equation} \citep[See, e.g.,][]{sunyaev01,vaughan01}. Here, $\tau$ is the optical depth, $\Delta t$ are the time lags, $k_{b}T_{e}$ is the plasma temperature and $E_{2}=16.0$ keV and $E_{1}=7.1$ keV are the energies of the photons in the broad energy bands we chose. We assume $\tau = 5$ and $k_{b}T_{e}=5$ keV as typical values for 4U 1636--53 \citep{fiocchi01,ming01,ming02}. From the size scales calculated with these values we can get the electron density using $n_{e}=\tau/(a\sigma_{T})$. See Table \ref{tab:sizeScalesDensityMedium}.

%
%
%
\begin{table}
\caption{Estimates of the scale size and density of the medium where $L_{b2}$, $L_{b}$, $L_{h}$, $L_{hHz}$, $L_{l}$ and $L_{u}$ QPOs of the NS-LMXB 4U 1636--53 are produced. Here, $c\Delta t$ gives the light travel time corresponding to the measured time lags (see Table \ref{tab:constantPhaseLagsValues}), $a$ is the size scale and $n_{e}$ is the electronic density of the medium (see Equation \ref{eq1} and text for the details of the calculations).}
\centering
\begin{tabular}{|c|c|c|c|}
\hline
QPO & $c\Delta t$ [km] & a [km] &  $n_{e}$ [$10^{20}$ $cm^{-3}$] \\
\hline
    $L_{b2}$ & $2610 \pm 630$ & $628.6 \pm 151.7$ & $0.0012 \pm 0.0003$ \\
    $L_{b}$ & $15 \pm 27$ & $3.6 \pm 6.5$ & $0.21 \pm 0.37$ \\
    $L_{h}$ & $240 \pm 30$ & $57.8 \pm 7.2$ & $0.013 \pm 0.002$ \\
    $L_{hHz}$ & $2.4 \pm 10.5$ & $0.58 \pm 2.53$ & $1.3 \pm 5.7$ \\
    $L_{l}$ & $6.3 \pm 0.6$ & $1.52 \pm 0.14$ & $0.50 \pm 0.05$ \\
    $L_{u}$ & $3.0 \pm 0.6$ & $0.72 \pm 0.14$ & $1.04 \pm 0.21$ \\
\hline
\hline
\end{tabular}
\label{tab:sizeScalesDensityMedium}
\end{table}

Adopting a Keplerian interpretation of the frequencies of the QPOs, we can infer the location where the QPOs themselves are produced. The higher the frequency, the closer to the compact object the QPOs originate. The values in Table \ref{tab:sizeScalesDensityMedium} suggest that a multi-component structure is needed for producing the time lags; at least two components, one for $L_{b}$, $L_{hHz}$, $L_{l}$ and $L_{u}$ with density $\sim10^{19-20}$ $\mathrm{cm}^{-3}$  and another for $L_{b2}$ and $L_{h}$ with density $\sim10^{17-18}$ $\mathrm{cm}^{-3}$. Notice the scale sizes for the first set are quite small (0.5 to 3.5 km), while for the latter are much bigger, 60 km and 630 km. It is interesting that the relations of fractional rms vs $\nu_{u}$ and Q (the quality factor, $\nu_{0}/FWHM$) vs $\nu_{u}$ are similar for $L_{b}$, $L_{hHz}$ and $L_{u}$ on one hand and for $L_{b2}$, $L_{h}$ and $L_{l}$ on the other \citep[see Figures \ref{fig:diagnostico1} and \ref{fig:diagnostico2} here and][for examples of these relations for other atoll sources.]{straaten01,diego01}. Again, this suggests that different mechanisms operate for these two sets of QPOs separately.

%
%

In recent years, models were developed to explain the $\sim 30$ sec soft time lags for frequencies $\geq 5\times10^{-4}$ Hz seen in AGNs \citep{zoghbi01,zoghbi02} in which the lags are due to reflection of coronal photons off the accretion disc. Although this scenario could explain the soft lags seen in the lower kHz QPO of a number of NS-LMXBs, the model faces difficulties to address other properties of this particular QPO such as the rms increase with energy in these systems (which, by the way is also difficult to reconcile with Compton down-scattering models; \citealt{nowak03,vaughan02,kaaret01}, for example. See \citealt{berger01,mendez06,gilfanov01} for discussions on these issues).

Following \cite{kotov01}, \cite{zoghbi01} suggested that while the hard lags seen for frequencies $\leq 5\times10^{-4}$ Hz in AGN are due to inward propagation of fluctuations in the disc, the soft lags seen above this frequency are due to reflection. The same reasoning could in principle be applied for the $L_{b2}$ and $L_{hHz}$ QPOs presented here, since their time lag-frequency spectrum show a loose resemblance to the lag-frequency spectrum in AGN.

\cite{deMarco01,deMarco02} studied the relation between black hole mass and soft X-ray time lags in AGNs and in ULX NGC 5408 X-1, suggesting that the relation holds all the way down to stellar-mass systems. In fact, in their Figure 9 \cite{deMarco02} show a schematic plot of the scaling relation for AGNs and the intermediate-mass black hole NGC 5408 X-1 where they also included the BHB GX 339-4 and the two NS-LMXBs 4U 1608--52 and 4U 1636--53. The relation of soft lags with mass appears to hold for neutron stars systems when for the latter one accounts only for the soft lags of the lower kHz QPO. This scaling relation has been interpreted as a signature of reverberation of the accretion disc in response to changes in the continuum.

The reverberation scenario \citep{zoghbi01,zoghbi02}, however, has difficulties to explain the lags of the other components whose lags are on average hard and do not scale with the mass.

Recently \cite{mendez08} measured the energy dependence of the phase lags of the high-frequency QPOs in four black hole candidates. For GRS 1915+105 the phase lag of the QPO at $35$ Hz is soft and the phase lag of the QPO at $67$ Hz is hard and inconsistent with the phase lags of the QPO at $35$ Hz. For IGR J17091--3624, the only QPO detected is at $67$ Hz and shows hard phase lags. For XTE J1550--564 the QPO at $180$ Hz shows hard phase lags, and the QPO at $280$ Hz shows lags consistent with zero or slightly positive. Finally, for GRO J1655--40 two QPOs were found, one at $300$ Hz with hard lags and the other at $450$ Hz with soft lags. 

It is tempting to try and compare those phase lags measured by \cite{mendez08} with the ones reported here in the bottom panels of Figure \ref{fig:phaseLagsVSfrequency}. It is possible that the same mechanism that produces the soft lags of the QPO at $35$ Hz and the hard lags of the QPO at $67$ Hz also applies to the lower and upper kHz QPOs of 4U 1636--36, since they also have opposites signs and are inconsistent with each other \cite[see][for additional discussion about this]{mendez08}; the hHz QPO of 4U 1636--36 could be somehow related to the QPO of the black hole candidates spanning the range $180-450$ Hz.

On the other hand, recently \cite{bult01} found a relation between the 401 Hz pulse frequency in the accreting millisecond pulsar SAX J1808.4-3658 and the upper kHz QPO detected in the range 300 to 700 Hz. They found that the amplitude of the pulse changes by a factor $\sim 2$ when the upper kHz QPO passes through 401 Hz. The findings of \cite{bult01} suggest that the upper kHz QPO originates from azimuthal motion at the inner edge of the disc (be it inhomogeneities or a geometric structure in the disc, like a bending wave moving with the orbital frequency). 


It is now reasonably well established that sustained accretion can spin up the neutron star in LMXBs to millisecond periods \citep[this is the classical Recycling Scenario; see][for a review]{bhattacharya01}. In this scenario, at least some of the LMXBs can be progenitors of radio millisecond pulsars. The discovery of the first millisecond X-ray pulsar, SAX J1808.4-3658 \citep{wijnands04}, and of the first black widow, B1957+20 \citep{fruchter01}, filled some gaps in the evolutionary path, making this scenario more plausible. 

Recently, \cite{papitto01} found what is understood as the swinging pulsar, considered the missing link between the X-ray accreting millisecond pulsars and the radio millisecond pulsars. The system in question, IGRJ18245--2452, switches from a X-ray to a radio pulsar due to intermittent accretion flow.

Both, SAX J1808.4-3658 and 4U 1636--53, are systems with neutron stars with millisecond spin period. However, while SAX J1808.4-3658 is an accreting millisecond X-ray pulsar, 4U 1636--53 does not exhibit X-ray pulsations, but X-ray bursts instead and by far most of LMXBs are not accreting X-ray pulsars. This is somewhat surprising since theory predicts that accreting neutron stars with magnetic field $B\sim 10^{8}$ G should be a X-ray pulsar because of the channeling of matter from the accretion flow through the magnetic field lines \citep{chakrabarty01}.


On the other hand, \cite{straaten03} studied a number of accreting millisecond X-ray pulsars and among them SAX J1808.4-3658. They found that the correlations between the timing features and colors of these pulsars are very similar to what is found in atoll sources, the subclass of NS-LMXBs to which 4U 1636--53 belongs to. Although SAX J1808.4-3658 shows a shifted correlation between the low-frequency components and $\nu_{u}$ by a factor of $1.5$, the mutual correlations between the low frequency components lie perfectly in what we observe in atoll sources.


Therefore, SAX J1808.4-3658 is appropriately classified as an atoll source and the absence of X-ray pulsations in 4U 1636--53 is not in contradiction with the picture described above and it is likely that the upper kHz QPO seen in this source reflects the same kind of azimuthal motion at the inner edge of the disc than SAX J1808.4-3658.

In the same line, \cite{bachetti01} and \cite{romanova01} can produce high frequency QPOs with 3D simulations of the accretion flow onto a magnetized neutron star, of particular interest for us in what they call magnetic boundary layer regime. 

Taking these two results together, the time lags of the kHz QPOs pair could constrain (to some extent) these models. We suggest that one possible assumption is that the time lags of the upper kHz QPO encode the properties of the medium at the magnetospheric radius (where the upper kHz QPO would be produced, at 16 to 21 km in our estimations\footnote{We assumed the neutron star mass in the range $M\sim 1.6 - 1.9$ $\mathrm{M_{\odot}}$, fixed the radius of the neutron star in $10$ km, fixed the magnetic field at $B\sim10^{8}$ G and assumed that the source accretes at a rate $0.03 - 0.06$ $\mathrm{L_{edd}}$.}, and where the scale size for the lags is $\sim 0.72$ km), while the time lags of the lower kHz QPOs encode the properties of the medium at the boundary layer and nearby (where the lower kHz QPO would be produced, near the surface of the neutron star, where the scale size for the lags is $\sim 1.52$ km). (For the size scales see $a$ in Table \ref{tab:sizeScalesDensityMedium}.) This idea would explain why the time lags of these two QPOs are inconsistent with each other and of opposite sign.

\subsection{Energy dependence}
\label{disEn}

While the frequency dependence of the phase lags relates to the geometry of the medium, the energy dependence relates to the physical conditions of the medium, like temperatures, densities and radiative processes. 

In the case of $L_{b2}$ we detect only a marginal trend of the phase lags with energy, with the phase lags becoming softer with increasing energy. $L_{b}$ also shows a marginal trend, but in this case the lags become harder with increasing energy. $L_{h}$ show one of the strongest trends with energy of all QPOs studied here. The phase lags are soft below 10 keV and hard above 10 keV). The case of $L_{hHz}$ is similar to $L_{b}$ and $L_{h}$: the phase lags begin soft at low energies and become hard at high energies. Regarding the lower and upper kHz QPOs we confirm the results of \cite{deAvellar01}: The phase lags of the lower kHz QPO are soft and decrease with energy whereas the lags of the upper kHz QPO are hard and do not depend upon energy.

The trends with energy show resemblance if we group together $L_{b}$, $L_{h}$, $L_{hHz}$ and $L_{u}$ on one hand and $L_{b2}$ and $L_{l}$ on the other. The lags of the first group become harder and those of the second group become softer as the energy increases. These trends may be a hint of a general mechanism behind the production of the lags of the QPOs of each group.

%
%

%
%

\cite{lee02} developed a model that can consistently explain the lag dependence upon energy seen in the lower kHz QPO of 4U 1608--52 with an up-scattering Comptonization model. In that model, the corona and disc temperatures oscillate coherently at the QPO frequency but the primary oscillations take place in the corona with the source of photons responding to the oscillations. \cite{lee02} constrained the size of the scattering plasma to 5 km for a fraction of $\eta\geq 0.5$ of up-scattered photons impinging back onto the source of soft photons using data of the lower kHz QPO at 830 Hz of the NS-LMXB 4U 1608--52, which shows time lags of a few $\mu$sec \citep{vaughan01,deAvellar01,barret06}. The size scales we estimated here are compatible with this view.

\cite{lee02} could predict not only the behaviour of the time lags of the lower kHz QPO with energy, but also  the fractional rms energy spectrum of this QPO. On the other hand, the fact that the oscillations cannot occur at two different frequencies simultaneously was used by \cite{deAvellar01} to explain the hard lags of the upper kHz QPO.

\cite{kumar01} developed a new model for the energy dependence of the time lags based on a thermal Comptonizing plasma oscillating at kHz frequencies. They tested their model against the measured time lags of the lower kHz QPO at $\sim 850$ Hz of the LMXB 4U 1608--52 by \cite{barret06}. \cite{kumar01} obtained soft lags only when there is a variation in the heating rate of the corona and a significant fraction of the photons impinges back into the photon source. They constrained the size of the Comptonizing region to $L=1.0$ km for $\eta=0.4$. Also, the model predicts consistently the fractional rms for energies $\leq 20$ keV. The hard lag-energy spectrum of the upper kHz QPO remains unexplained by their model. It is interesting to note that within the scope of this model, if one takes into account that we can group the lag-energy spectrum of $L_{l}$ and $L_{b2}$ together (the only other soft lags we found), the same mechanism could apply. However, the lag-energy spectrum of the upper kHz QPO and of the others is beyond this model up to now.

The model in \cite{kumar01} can also explain the increase of the fractional rms with energy satisfactorily, although cannot explain the flattening at energies above 10 keV seen in the data \citep[see, for example,][for 4U 1608--52 and 4U 1636--53, respectively]{kumar01,diego01}.

More recently \cite{peille01} performed an extensive study of the spectral-timing properties of the lower and upper kHz QPOs in the source 4U 1728--34. They also performed the first spectral deconvolution of the covariance spectra (which is equivalent to the rms spectrum if the variations are strongly correlated across all energies) of both kHz QPOs.

They found that the QPO spectrum is compatible with the one of a Comptonized black body with temperature higher than the temperature of the continuum, which implies that a more compact inner region of the boundary layer is producing the QPO. This Comptonized black body seems to be indispensable, since a substitution of this component for a power law does not fit the data well. Another important findings are that the lag-energy spectrum of the lower kHz and the upper kHz QPO are systematically different in the same way we have found in \cite{deAvellar01} and that they also depend weakly on frequency.    

An interesting scenario then arises: if the lags of the upper kHz QPO are dominated by reverberation this would imply the origin of this QPO as simple variations in luminosity due to variation in the accretion rate on the boundary layer. These variations in the accretion rate would occur at the innermost regions of the accretion disc, an idea supported by the findings of \cite{bult01}. In this scenario, we see the upper kHz QPO signal as a response of the boundary layer to these variation. Evidence that broader and lower frequency noise components are also generated by the same kind of mechanism \citep{uttley01,uttley02} would then be corroborated by our results based on the general shapes of the lag-energy spectrum, except for $L_{b2}$ which is similar to the lower kHz QPO $L_{l}$.

The more coherent signal of the lower kHz QPO may have its origin in a more compact region of the boundary layer itself due to internal oscillations in the heating rate of the boundary layer \citep[as in ][for example]{lee02,kumar01}. This is corroborated in part by the recent work of \cite{cackett01}. He found that the lag-energy spectrum of the lower kHz QPO of 4U 1608--52 cannot be due to only reverberation. His modelling of the lag-energy spectrum of the lower kHz QPO of that source showed that the lags would be expected to increase above $\sim 8$ keV, in contradiction with the observations. He argue that another physical mechanism is required to produce  the observed lag.

Our results for the lag energy spectrum of the QPOs seem to point also in that direction.

If extended to include all the other QPOs, these models provide an opportunity to study the dynamic and physical conditions of the Comptonising corona in neutron-star low-mass X-ray binaries.

\section*{Acknowledgments}

MGBA acknowledges the financial support from FAPESP 2011/23996-9 and from the FAPESP Thematic Project 2013/26258-4. MGBA is also grateful to the Kapteyn Astronomical Institute for the hospitality and to R. Misra, M. Linares, P. Bult and T. Strohmayer for the useful comments. DA acknowledges support from the Royal Society. The authors are grateful to the unknown referee for his/her meticulous review of this paper whose comments significantly increased the value of our work.

\clearpage

%
%
%


\clearpage


\label{lastpage}


\begin{thebibliography}{99}
\bibitem[\protect\citeauthoryear{Alston et al.}{2015}]{alston01}
Alston W. N., Parker M. L., Markevi\v ci\=ut\.e J., Fabian A. C., Middleton M., Lohfink A., Kara E., Pinto C., 2015, MNRAS, 449, 467
\bibitem[\protect\citeauthoryear{Altamirano et al.}{2008}]{diego01}
Altamirano, D., van der Klis, M., M\'endez, M., Jonker, P., Klein-Wolt, M., Lewin, W. H. G. 2008, ApJ, 685, 436
\bibitem[\protect\citeauthoryear{Altamirano \& M\'endez}{2015}]{diego02}
Altamirano, D., M\'endez, M., MNRAS, 2015, 449, 4027
\bibitem[\protect\citeauthoryear{Bachetti et al.}{2010}]{bachetti01}
Bachetti, M., Romanova, M., Kulkarni, A., Burderi, L., di Salvo, T. 2010, MNRAS, 403, 1193
\bibitem[\protect\citeauthoryear{Bhattacharya \& van den Heuvel}{1991}]{bhattacharya01} 
Bhattacharya, D., van den Heuvel, E.~P.~J., 1991, PhR, 203, 1
\bibitem[\protect\citeauthoryear{Barret}{2013}]{barret06}
Barret, D., 2013, ApJ, 770, 9
\bibitem[\protect\citeauthoryear{Belloni et al}{2007}]{belloni01} 
{Belloni}, T., Homan, J., Motta, S., Ratti, E. and M{\'e}ndez, M. 2007, MNRAS, 379, 247 
\bibitem[\protect\citeauthoryear{Berger et al.}{1996}]{berger01}
Berger, M., van der Klis, M., van Paradijs, J., Lewin, W. H. G., Lamb, F., Vaughan, B., Kuulkers, E., Augusteijn, T., Zhang, W., Marshall, F. E., Swank, J. H., Lapidus, I., Lochner, J. C., Strohmayer, T. E. 1996, ApJL, 469, L13
\bibitem[\protect\citeauthoryear{B\"ock et al.}{2011}]{bock01}
B\"ock M., Grinberg V., Pottschmidt K., Hanke M., Nowak M. A., Markoff S. B., Uttley P., Rodriguez J., Pooley G. G., Suchy S., Rothschild R. E., Wilms J., 2011, A\& A, 533 A8
\bibitem[\protect\citeauthoryear{Bradt, Rothschild \& Swank}{1993}]{bradt01}
Bradt, H. V., Rothschild, R. E., Swank, J. H., 1993, A\& AS, 97, 355
\bibitem[\protect\citeauthoryear{Bult \& van der Klis}{2015}]{bult01} 
Bult, P., van der Klis, M. 2015, ApJL, 798, Issue 2, L29 
\bibitem[\protect\citeauthoryear{Bussard et al}{1988}]{bussard01}
Bussard, R. W., Weisskopf, M. C., Elsner, R. F., Shibazaki, N. 1988, ApJ, 327, 284.
\bibitem[\protect\citeauthoryear{Cackett}{2016}]{cackett01} 
Cackett, E. M., 2016, arXiv:1601.07849v1, {\it submitted to ApJ}.
\bibitem[\protect\citeauthoryear{Cassatella et al.}{2012}]{cassatella01} 
Cassatella, P. and Uttley, P. and Wilms, J. and Poutanen, J. 2012, MNRAS, 422, 2407
\bibitem[\protect\citeauthoryear{Chakrabarty}{2005}]{chakrabarty01} 
{Chakrabarty}, D. 2005, Astronomical Society of the Pacific Conference Series, 328, 279, astro-ph/0408004
\bibitem[\protect\citeauthoryear{Cumming et al.}{2001}]{cumming01} 
{Cumming}, A., {Zweibel}, E., {Bildsten}, L. 2001, ApJ, 557, 958
\bibitem[\protect\citeauthoryear{de Avellar et al.}{2013}]{deAvellar01} 
de Avellar, M. G. B., M\'endez, M., Sanna, A., Horvath, J. E. 2013, MNRAS, 433, 3453
\bibitem[\protect\citeauthoryear{de Avellar et al.}{2014}]{deAvellar02} 
de Avellar, M. G. B.; M\'endez, M.; Altamirano, D.; Sanna, A.; Zhang, G., The X-ray Universe 2014, edited by Jan-Uwe Ness.
\bibitem[\protect\citeauthoryear{De Marco et al.}{2013a}]{deMarco01}
De Marco, B., Ponti, G., Cappi, M., Dadina, M., Uttley, P., Cackett, E. M., Fabian, A. C., Miniutti, G. 2013, MNRAS, 431, 2441
\bibitem[\protect\citeauthoryear{De Marco et al.}{2013b}]{deMarco02}
De Marco, B.; Ponti, G.; Miniutti, G.; Belloni, T.; Cappi, M.; Dadina, M.; Muñoz-Darias, T. 2013, MNRAS, 436, 3782
\bibitem[\protect\citeauthoryear{Di Salvo, M\'endez \& van der Klis}{2003}]{disalvo01}
Di Salvo, T., M{\'e}ndez, M., van der Klis, M. 2003, A\& A, 406, 177
\bibitem[\protect\citeauthoryear{Dieters, Vaughan, Kuulkers \& van der Klis}{1997}]{dieters01}
Dieters, S., Vaughan, B. A., Kuulkers, E. \& van der Klis, M. 1997, AA, in preparation
\bibitem[\protect\citeauthoryear{Done et al}{2007}]{done01}
Done, C., Gierli\'nski, M., Kubota, A. 2007, A\& ARv, 15, issue 1, 1
\bibitem[\protect\citeauthoryear{Esin et al}{1997}]{esin01}
Esin, A.~A., McClintock, J.~E. and Narayan, R. 1997, ApJ, 489, 865
\bibitem[\protect\citeauthoryear{Falanga \& Titarchuk}{2007}]{falanga01}
Falanga, M., Titarchuk, L. 2007, ApJ, 661, 1084
\bibitem[\protect\citeauthoryear{Fiocchi et al}{2006}]{fiocchi01}
Fiocchi, M. and Bazzano, A. and Ubertini, P. and Jean, P. 2006, ApJ, 651, 416
\bibitem[\protect\citeauthoryear{Ford et al.}{1999}]{ford01} 
Ford, E. C., van der Klis, M., M{\'e}ndez, M., van Paradijs, J., Kaaret, P. 1999, ApJ, 512, L31 
\bibitem[\protect\citeauthoryear{Frank, King \& Raine}{2002}]{frank01}
Frank, J., King, A., Raine, D., Cambridge University Press, February 2002
\bibitem[\protect\citeauthoryear{Fruchter, Stinebring \& Taylor}{1988}]{fruchter01}
Fruchter, A.~S., Stinebring, D.~R., Taylor, J.~H., 1988, Nature, 333, 237
\bibitem[\protect\citeauthoryear{Gilfanov, Revnivtsev, \& Molkov}{2003}]{gilfanov01} 
Gilfanov, M., Revnivtsev, M., Molkov, S. 2003, A\&A, 410, 217 
\bibitem[\protect\citeauthoryear{Grinberg et al}{2014}]{grinberg01}
Grinberg V., Pottschmidt K., B\"ock M., Schmid C., Nowak M. A., Uttley P., Tomsick J. A., Rodriguez J., Hell N., Markowitz A., Bodaghee A., Cadolle Bel M., Rothschild R. E., Wilms J., 2014, A\& A, 565, A1
\bibitem[\protect\citeauthoryear{Hua, Kazanas, \& Titarchuk}{1997}]{hua01} 
Hua, X. M., Kazanas, D., Titarchuk, L. 1997, ApJ, 482, L57 
\bibitem[\protect\citeauthoryear{Jahoda et al.}{2006}]{jahoda01}
Jahoda, K., Markwardt, C. B., Radeva, Y., Rots, A. H., Stark, M. J., Swank, J. H., Strohmayers, T. E., Zhang, W., 2006, ApJS, 163, 401
\bibitem[\protect\citeauthoryear{Kaaret et al.}{1999}]{kaaret01}
Kaaret, P., Piraino, S., Ford, E., Santangelo, A. 1999, ApJ, 514, L31
\bibitem[\protect\citeauthoryear{Kolgomorov}{1933}]{kolgomorov01} 
A. Kolmogorov, Giornale dell' Instituto Italiano degli Attuari 4 (1933), p. 83.
\bibitem[\protect\citeauthoryear{Kotov et al.}{2001}]{kotov01} 
Kotov, O., Churazov, E., Gilfanov, M., 2001, MNRAS, 327, 799
\bibitem[\protect\citeauthoryear{Kumar \& Misra}{2014}]{kumar01}
Kumar, N., Misra, R., MNRAS, 2014, 445, 2818
\bibitem[\protect\citeauthoryear{Kuulkers et al}{1994}]{kuulkers01}
Kuulkers, E., van der Klis, M., Oosterbroek, T., Asai, K., Dotani, T., van Paradijs, J., Lewin, W.~H.~G., A\& A, 289, 795
\bibitem[\protect\citeauthoryear{Lee \& Miller}{1998}]{lee01}
Lee, H. C., Miller, G. S. 1998, MNRAS, 299, 479
\bibitem[\protect\citeauthoryear{Lee, Misra \& Taam}{2001}]{lee02}
Lee, H. C., Misra, R., Taam, R. E. 2001, ApJ, 549, L229
\bibitem[\protect\citeauthoryear{Lei et al}{2014}]{lei01}
Lei, Y., Zhang, S., Qu, J., Yuan, H., Wang, Y., Dong Y., Zhang, H., Li, Z., Zhang, C., and Zhao, Y., 2014, ApJ, 147, 67
\bibitem[\protect\citeauthoryear{Linares}{2009}]{linaresThesis}
Linares, M., 2009, Ph.D. Thesis
\bibitem[\protect\citeauthoryear{Mauche}{2002}]{mauche01}
Mauche, C. W. 2002, ApJ, 580, 423
\bibitem[\protect\citeauthoryear{Melo et al.}{2009}]{melo01}
{Melo}, I., {Tom{\'a}{\v s}ik}, B., {Torrieri}, G., {Vogel}, S., {Bleicher}, M., {Kor{\'o}ny}, S., {Gintner}, M. 2009, PhRvC, 80, 024904
\bibitem[\protect\citeauthoryear{M\'endez}{2006}]{mendez04}
M\'endez, M. 2006, MNRAS, 371, 1925
\bibitem[\protect\citeauthoryear{M\'endez et al.}{2013}]{mendez08}
M\'endez, M., Altamirano, D., Belloni, T., Sanna, A. 2013, MNRAS, 435, 2132
\bibitem[\protect\citeauthoryear{M\'endez, van der Klis \& Ford}{2001}]{mendez06}
M{\'e}ndez, M., van der Klis, M., Ford, E. C. 2001, ApJ, 561, 1016
\bibitem[\protect\citeauthoryear{M\'endez et al.}{1999}]{mendez03}
M\'endez, M., van der Klis, M., Ford, E. C., Wijnands, R., van Paradijs, J. 1999, ApJ, 511, L49
\bibitem[\protect\citeauthoryear{Meyer-Hofmeister et al.}{2005}]{meyerHofmeister01}
Meyer-Hofmeister, E., Liu, B.~F. and Meyer, F. 2005, A\& A, 432, 181
\bibitem[\protect\citeauthoryear{Miller, Lamb \& Psaltis}{1998}]{miller01}
Miller, M. C., Lamb, F. K., Psaltis, D. 1998, ApJ, 508, 791
\bibitem[\protect\citeauthoryear{Ming, M\'endez \& Altamirano}{2014}]{ming01}
Lyu, M. and M{\'e}ndez, M. and Altamirano, D. 2014, MNRAS, 445, 3659
\bibitem[\protect\citeauthoryear{Ming et al.}{2015}]{ming02}
Lyu, M. and M{\'e}ndez, M. and Zhang, G. and Keek, L. 2015, MNRAS, 454, 541
\bibitem[\protect\citeauthoryear{Mitsuda \& Dotani}{1989}]{mitsuda01}
Mitsuda, K., \& Dotani, T. 1989, PASJ, 41, 557
\bibitem[\protect\citeauthoryear{Miyamoto et al.}{1988}]{miyamoto01} 
Miyamoto, S., Kitamoto, S., Mitsuda, K., Dotani, T. 1988, Natur, 336, 450 
\bibitem[\protect\citeauthoryear{Nowak et al.}{1999}]{nowak01}
Nowak, M. A., Vaughan, B. A., Wilms, J., Dove, J. B., Begelman, M. C. 1999, ApJ, 510, 874
\bibitem[\protect\citeauthoryear{Nowak et al.}{1996}]{nowak03} 
Nowak, M. A., Vaughan, B. A. 1996, MNRAS, 280, 227 
\bibitem[\protect\citeauthoryear{Papitto et al.}{2013}]{papitto01}
{Papitto}, A., {Ferrigno}, C., {Bozzo}, E., {Rea}, N., {Pavan}, L., {Burderi}, L., {Burgay}, M., {Campana}, S., {di Salvo}, T., {Falanga}, M., {Filipovi{\'c}}, M.~D., {Freire}, P.~C.~C., {Hessels}, J.~W.~T., {Possenti}, A., {Ransom}, S.~M., {Riggio}, A., {Romano}, P., {Sarkissian}, J.~M., {Stairs}, I.~H., {Stella}, L., {Torres}, D.~F., {Wieringa}, M.~H., {Wong}, G.~F. 2013, Nature, 501, 517
\bibitem[\protect\citeauthoryear{Payne}{1980}]{payne01}
Payne, D. G. 1980, ApJ, 237, 951
\bibitem[\protect\citeauthoryear{Peille, Barret \& Uttley}{2015}]{peille01}
Peille, P., Barret, D., Uttley, P., ApJ, 2015, 811, 109
\bibitem[\protect\citeauthoryear{Psaltis, Belloni \& van der Klis}{1999}]{psaltis01}
Psaltis, D., Belloni, T., van der Klis, M., 1999, ApJ, 520, 262
\bibitem[\protect\citeauthoryear{Reig et al.}{2004}]{reig01}
{Reig}, P., {van Straaten}, S., {van der Klis}, M., 2004, ApJ 602, 918
\bibitem[\protect\citeauthoryear{Romanova \& Kulkarni}{2009}]{romanova01}
Romanova, M., Kulkarni, A. 2009, MNRAS, 398, 1105
\bibitem[\protect\citeauthoryear{Sanna et al.}{2012}]{sanna02}
Sanna, A., M\'endez, M., Belloni, T., Altamirano, D. 2012, MNRAS, 424, 2936
\bibitem[\protect\citeauthoryear{Stella \& Vietri}{1999}]{stella01}
Stella, L., Vietri, M. 1999, PRL, 82, 17
\bibitem[\protect\citeauthoryear{Stollman et al}{1987}]{stollman01}
Stollman, G. M., van Paradijs, J., Hasinger, G., Lewin, W. H. G., van der Klis, M. 1987, MNRAS, 227, 7
\bibitem[\protect\citeauthoryear{Sunyaev \& Titarchuk}{1980}]{sunyaev01}
Sunyaev, R. A., Titarchuk, L. G. 1980, A\& A, 86, 121
\bibitem[\protect\citeauthoryear{Uttley}{2004}]{uttley01}
Uttley, P., MNRAS, 347, L61
\bibitem[\protect\citeauthoryear{Uttley et al}{2011}]{uttley02}
Uttley, P., Wilkinson, T., Cassatella, P., Wilms, J., Pottschmidt, K., Hanke, M. and B{\"o}ck, M., MNRAS, 414, L60
\bibitem[\protect\citeauthoryear{van der Klis}{2006}]{vanderklis01}
van der Klis, M. 2006, in Lewin, W., van der Klis, M., eds., Cambridge Astrophysics Ser. Vol. 39, Compact stellar X-ray sources, Cambridge University Press, Cambridge  p. 39 
\bibitem[\protect\citeauthoryear{van der Klis et al}{1987}]{vanderklis02}
van der Klis, M., Hasinger, G., Stella, L., Langmeier, A., van Paradijs, J., and Lewin, W. H. G., 1987, ApJL, 319, L13
\bibitem[\protect\citeauthoryear{van Straaten et al.}{2002}]{straaten02}
van Straaten, S., van der Klis, M., di Salvo, Tiziana, Belloni, T., 2002, ApJ, 568, 912
\bibitem[\protect\citeauthoryear{van Straaten, van der Klis, \& M\'endez}{2003}]{straaten01}
van Straaten, S., van der Klis, M., M\'endez, M. 2003, ApJ, 596, 1155
\bibitem[\protect\citeauthoryear{van Straaten, van der Klis, \& Wijnands}{2005}]{straaten03}
van Straaten, S., van der Klis, M., Wijnands, R. 2005, ApJ, 619, 455
\bibitem[\protect\citeauthoryear{Vaughan \& Nowak}{1997a}]{vaughan03}
Vaughan, B. A., Nowak M. A. 1997a, ApJ, 474, L43
\bibitem[\protect\citeauthoryear{Vaughan et al.}{1994}]{vaughan04}
Vaughan, B. A., van der Klis, M., Lewin, W. H. G., Wijers, R. A. M. J., van Paradijs, J., Dotani, T., Mitsuda, K. 1994, ApJ, 421, 738
\bibitem[\protect\citeauthoryear{Vaughan et al.}{1997b}]{vaughan01}
Vaughan, B. A., van der Klis, M., M\'endez, M., van Paradijs, J., Wijnands, R. A. D., Lewin, W. H. G., Lamb, F. K., Psaltis, D., Kuulkers, E., Oosterbroek, T. 1997b, ApJ, 483, L115
\bibitem[\protect\citeauthoryear{Vaughan et al.}{1998}]{vaughan02}
Vaughan, B. A., van der Klis, M., M\'endez, M., van Paradijs, J., Wijnands, R. A. D., Lewin, W. H. G., Lamb, F. K., Psaltis, D., Kuulkers, E., Oosterbroek, T. 1998, ApJ, 509, L145
\bibitem[\protect\citeauthoryear{Warner \& Woudt}{2002}]{warner01} 
Warner, B., Woudt, P. A. 2002, MNRAS, 335, 84
\bibitem[\protect\citeauthoryear{Wijers, van Paradijs \& Lewin}{1987}]{wijers01} 
Wijers, R. A. M. J., van Paradijs, J. \& Lewin, W. H. G. 1987, MNRAS, 228, 17
\bibitem[\protect\citeauthoryear{Wijnands et al.}{1997}]{wijnands01} 
Wijnands, R. A. D., van der Klis, M., van Paradijs, J., Lewin, W. H. G., Lamb, F. K., Vaughan, B., Kuulkers, E. 1997, ApJ, 479, L141 
\bibitem[\protect\citeauthoryear{Wijnands \& van der Klis}{1999}]{wijnands02} 
Wijnands, R. A. D., van der Klis, M., 1999, ApJ, 514, 939
\bibitem[\protect\citeauthoryear{Wijnands et al.}{1997}]{wijnands03} 
Wijnands, R. A. D., van der Klis, M., Kuulkers, E., Asai, K., Hasinger, G., 1997, A\& A, 323, 399
\bibitem[\protect\citeauthoryear{Wijnands et al.}{1997}]{wijnands04} 
{Wijnands}, R. and {van der Klis}, M., 1998, Nature, 394, 344
\bibitem[\protect\citeauthoryear{Zhang et al.}{2013}]{guobao01}
Zhang, G., M\'endez, M., Belloni, T. M.; Homan, J., 2013, MNRAS, 436, 2276
\bibitem[\protect\citeauthoryear{Zoghbi et al.}{2010}]{zoghbi01}
Zoghbi, A., Fabian, A. C., Uttley, P., Miniutti, G., Gallo, L. C., Reynolds, C. S., Miller, J. M., Ponti, G. 2010, MNRAS, 401, 2419
\bibitem[\protect\citeauthoryear{Zoghbi, Uttley \& Fabian}{2011}]{zoghbi02}
Zoghbi, A., Uttley, P., Fabian, A. C. 2011, MNRAS, 412, 59
\end{thebibliography}
\end{document}